\documentclass[aps,prc,groupedaddress,twocolumn,showpacs,amsmath,amssymb,floatfix]{revtex4}
\usepackage{graphics}
\usepackage{dcolumn}
\usepackage{longtable}

\begin{document}

\title{Alternate Solution of the $\alpha$-Potential Mystery}
\author{V.~Avrigeanu} \email{vlad.avrigeanu@nipne.ro}
\author{M.~Avrigeanu} \email{marilena.avrigeanu@nipne.ro}
\affiliation{Horia Hulubei National Institute for Physics and Nuclear Engineering, P.O. Box MG-6, 077125 Bucharest-Magurele, Romania}

\begin{abstract}
A consistent set of statistical-model input parameters, validated by analysis of various independent data, makes possible the assessment of an $\alpha$-particle optical model potential [Phys. Rev. C {\bf 90}, 044612 (2014)] also for nucleon-induced $\alpha$-emission within $A$$\sim$60 mass-number range.
The advantage of recent data for low-lying states feeding is taken as well.
Consideration of additional reaction channels leading to increase of the $\alpha$-emission beyond the statistical predictions has concerned the pickup direct interaction and Giant Quadrupole Resonance similar features. 
\end{abstract}

\maketitle

{\it Introduction.} -- The $\alpha$-particle interaction with nuclei as well as the corresponding optical model potential (OMP) were of special interest from nuclear-physics earliest days. 
The widely-used phenomenological OMP parameters have been derived from analysis of either elastic-scattering, which is ruled out below the Coulomb barrier $B$, or $\alpha$-induced reaction data. 
They are then used to describe also the $\alpha$--emission from excited nuclei, in terms of the statistical Hauser-Feshbach (HF) \cite{wh52p} and pre-equilibrium emission (PE) \cite{eg92p} models. 
However, there are also various assumptions and parameters of these models in addition to the $\alpha$-nucleus potential. 
Thus, a definite conclusion on $\alpha$-potential may become possible only using HF+PE consistent parameter sets already validated by analysis of other independent data distinct from either $\alpha$-induced reaction or especially $\alpha$--emission data. 

Moreover, there is a so-called $\alpha$-potential mystery of the account at once of both absorption and emission of $\alpha$-particles in nuclear reactions \cite{tr13p}, of equal interest for nuclear astrophysics and fusion technology.
It was referred by Rauscher \cite{tr13p} to competition of the Coulomb excitation (CE) with the compound nucleus (CN) formation in $\alpha$-induced reactions below $B$, while the former does not affect the $\alpha$--emission. 
Because the corresponding partial waves and integration radii may provide evidence for distinct account of CE cross section and OM total-reaction cross section \cite{va16ap}, an alternate solution may also concern additional reaction channels but for $\alpha$--emission. Thus, further consideration should be given to eventual role of the pickup direct interaction (DI) leading to increase of the $\alpha$-emission beyond the HF+PE predictions. 
Besides, a reaction cross-section enhancement that could be related to the position of a giant quadrupole resonance (GQR) is also worthy of note \cite{ma06p}. 

In fact, a former search for new physics in potentials to describe nuclear de-excitation assumed that particle evaporation occurs from a transient nuclear stratosphere of the emitter nucleus (\cite{ma06p} and Refs. therein). 
Recent definite conclusion on incident $\alpha$-particle OMP \cite{va14p} made feasible the analysis of a possible difference between the OMPs describing either $\alpha$-particle elastic scattering and reactions or $\alpha$-emission from excited nuclei \cite{va94p}. 
On the other hand, OMP \cite{va14p} validation for $\alpha$-emission in low-energy proton-induced reactions on Zn isotopes was related to a surface character \cite{va15p}, at variance with the fast-neutron induced reactions on Zr isotopes \cite{va17p}.
Analysis of $\alpha$--emission from nuclei excited in reactions induced by either neutrons or low-energy protons became thus of interest while there are quite useful recent data of low-lying states feeding \cite{gz10ap,gz10p,yg14p,zw15p,tk18p,hb19p} for $A$$\sim$60 nuclei.  

Consequently, it is of interest to find out also under these conditions if the $\alpha$-emission could be described by the same OMP \cite{va14p} which led to a rather good account of $\alpha$-induced reaction data \cite{va14p,ma09p,ma10p,va16p,va19p}. 
It provided in the meantime better results also within large-scale nuclear-data evaluation \cite{ajk15p}, being adopted as default option within the world-wide used code TALYS \cite{TALYSp}. 
Nevertheless, use of a consistent model parameter set, with no empirical rescaling factors of the $\gamma$ and/or neutron widths, is essential to avoid compensation effects. 
Hence the main HF, PE, and DI assumptions and parameters as well as independent data analysis to support them are given in the Supplemental Material  \cite{SM2020p}. 

\begin{figure*} 
\resizebox{1.8\columnwidth}{!}{\includegraphics{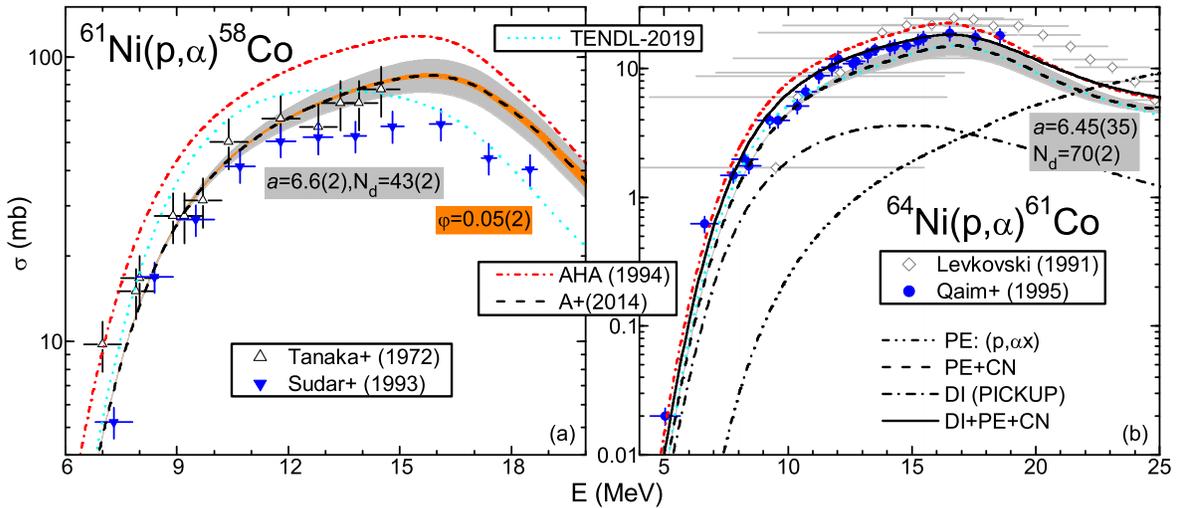}}
\caption{\label{Fig:Ni614pa} (Color online) Comparison of (a,b) measured $(p,\alpha)$  excitation functions on $^{61,64}$Ni \cite{exforp}, evaluated (short-dashed curves) \cite{TENDLp}, and PE+CN calculated values using the $\alpha$-particle OMPs of Refs. \cite{va14p} (dashed curves) and \cite{va94p} (short-dash-dotted curves), and (b) total PE $\alpha$-emission (dash-dot-dotted curve), DI pickup (dash-dotted curve), and sum (solid curve); uncertainty bands correspond to error bars of NLD parameter $a$ (with uncertainties in units of the last digit \cite{SM2020p}) and low-lying level cumulative number $N_d$ of residual nuclei (gray band), and (a) for GDH $\alpha$-particle pre-formation probability $\varphi$ (orange band).}
\end{figure*}

{\it DI pickup in addition to PE+CN $\alpha$-emission.} -- 
The due consideration of DI role for $\alpha$-emission in nucleon--induced reactions is obviously leading to an increase beyond the PE+HF predictions.
Actually, Qaim {\it et al.} \cite{smq95p} extended at incident energies $\leq$15 MeV the conclusion of '90s that the pickup instead of knockout has the main DI contribution to low-lying states in $(p,\alpha)$ and $(n,\alpha)$ reactions (\cite{eg92p,eg89p,eg91p} and Refs. therein). 
However, Qaim {\it et al.} normalized the semimicroscopic pickup contribution to their $^{64}$N$(p,\alpha)$$^{61}$Co data at 15 MeV, with results that depend notably on the former PE+HF component \cite{SM2020p}. 
Hence, an increased predictive power of the model results demands the primary use of  spectroscopic data.

In the present work, the pickup contribution to $(p,\alpha)$ and $(n,\alpha)$ reactions was determined within the distorted wave Born approximation (DWBA) method using the code FRESCO \cite{FRESCOp} and the same OMPs as within HF+PE analysis.
One--step reaction was considered through the pickup of $^3$H and $^3$He clusters, respectively. 
Moreover, the "spectator model" \cite{smits76p,smits79p} was involved, where the two transferred either neutrons or protons in $(p,\alpha)$ and $(n,\alpha)$ reactions, respectively, are coupled to zero angular momentum acting as spectators, while the transferred orbital ($L$) and total ($J$) angular momenta are given by the third, unpaired nucleon of the transferred cluster.
The prior form distorted--wave transition amplitudes, and the finite--range interaction were considered, with more details in \cite{SM2020p}. 
Finally, 19 states \cite{smits79p,plj78p,ensdfp}, until the excitation energy of $\sim$5 MeV, were involved for assessment of the pickup excitation function in Fig.~\ref{Fig:Ni614pa}(b).

First, within 5--7 MeV above the effective $(p,\alpha)$ reaction threshold, the DI pickup component is around half the CN one but an order of magnitude larger than PE contribution. 
The major role of DI pickup on low-lying residual states, at lower incident energies, is thus proved. 
The PE component corresponds to the Geometry-Dependent Hybrid (GDH) model \cite{mb83p}, generalized through inclusion of the angular-momentum and parity conservation \cite{ma90p} and the phenomenological knockout model of $\alpha$--emission based on a pre-formation probability $\varphi$ \cite{eg92p}. 
It increases with the incident energy, becoming dominant at proton energies $>$15 MeV. 

Second, the DI pickup inclusion is providing a suitable description of the measured data at all energies. 
This point is particularly notable below 10 MeV, where there is no effect of nuclear level density (NLD) parameters \cite{SM2020p} as shown by the related uncertainty bands.

Third, there is an obvious data agreement, within several MeV above the effective $(p,\alpha)$ reaction threshold, for the CN component given by the $\alpha$-particle OMP of Ref. \cite{va94p}. 
This fact may explain the former results found in this energy range but without DI account \cite{va94p}. 

The case of the $(p,\alpha)$ reaction on $^{61}$Ni is also shown in Fig.~\ref{Fig:Ni614pa}(a) making use of the CN isotopic effect of cross--section decreasing with the isotope mass increase \cite{nim77p}. 
The lower similar dependence of the DI processes leads to an increased sensitivity of the $(p,\alpha)$ reaction on heavier isotope $^{64}$Ni to the pickup contribution. 
Moreover, the larger cross sections for this case underlines the difference between predictions of the $\alpha$-particle OMPs \cite{va14p,va94p}, and thus support the former one. 

{\it DI pickup contribution in $\alpha$--emission spectra.} -- 
An additional check of DI pickup component of the $\alpha$--emission in neutron--induced reactions on $^{59}$Co is worthwhile, as follows. 
First, there is a large data basis of total $\alpha$--emission as well as $(n,\alpha)$ reaction cross sections over three decades. 
Second, $^{59}$Co may be considered as a benchmark because of its medium  asymmetry parameter $(N-Z)/A$ and corresponding isotopic effect. 
Third, there are measured at the incident energy of $\sim$14 MeV valuable $\alpha$--emission angular distributions and spectra \cite{rf86p,smg96p,k99p} that are quite important for DI+PE analysis. 

\begin{figure} 
\resizebox{0.85\columnwidth}{!}{\includegraphics{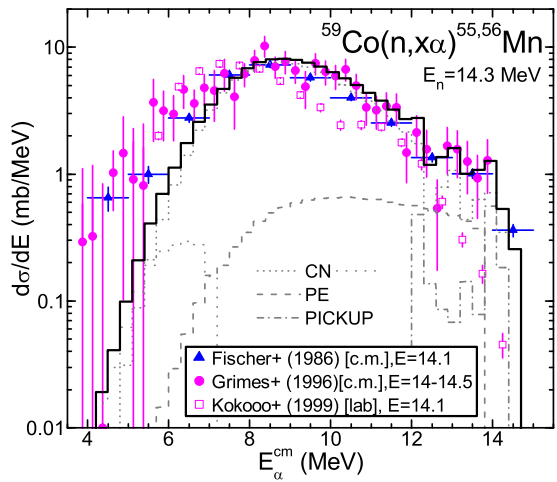}}
\caption{\label{Fig:Co59naS} Comparison of measured $\alpha$-emission spectra from 14--14.5 MeV neutron--induced reactions on $^{59}$Co \cite{rf86p,smg96p,k99p}, and calculated  DI pickup (dash-dotted curve), PE (dashed curves), CN first-- (short-dash curve) and second--emission (dotted curve) components at 14.3 MeV, and their sum (solid curve).}
\end{figure}

Actually, only a tentative attempt was possible in this work, using neutron spectroscopic factors from the $^{55}$Mn$(d,p)^{56}$Mn reaction analysis \cite{NDS56p,comfortp} and the spectator protons pair as picked from 1$f_{7/2}$ subshell  \cite{SM2020p}. 
Then, 27 excited states with well-known $J^{\pi}$ and transferred $L$ \cite{ensdfp,ripl3p}, until 2.088 MeV excitation energy, were considered in the pickup cross--section calculation.
The good agreement proved by the analysis of angular distributions for the $\alpha$-energy bins, within center-of-mass system (c.m.), of 6-10 MeV, 10-12 MeV, and 12-14 MeV (Fig. 26 of \cite{SM2020p}), is completed hereafter by $\alpha$-emission spectra and excitation function discussion. 

\begin{figure} 
\resizebox{0.85\columnwidth}{!}{\includegraphics{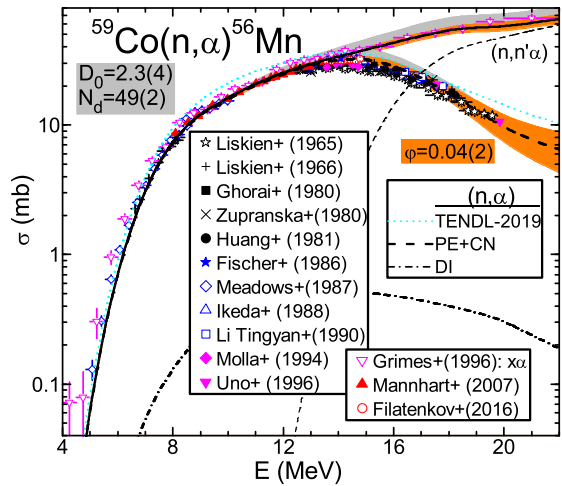}}
\caption{\label{Fig:Co59na} As Fig.~\ref{Fig:Ni614pa} but for $(n,\alpha)$ reaction on $^{59}$Co \cite{exforp} and uncertainty gray band for error bars of $s$-wave neutron-resonance spacing $D_0^{\it exp}$ and $N_d$ of residual nucleus $^{56}$Mn (Table I \cite{SM2020p}), and additionally calculated cross sections of $(n,n'\alpha)$ reaction (thin dashed curve) and total $\alpha$-emission (solid curve).}
\end{figure}

Nevertheless, the results in Fig.~\ref{Fig:Co59naS} support the present attempt of the DI pickup contribution on the lowest-lying states of the odd--odd residual nucleus $^{56}$Mn.
The data agreement could be increased by taking into account that data transformation from laboratory system (lab) to c.m. may provide, as shown for $^{56}$Fe \cite{rf84p}, a shift of up to 1 MeV to higher emission energies. 

\begin{figure*} 
\resizebox{1.8\columnwidth}{!}{\includegraphics{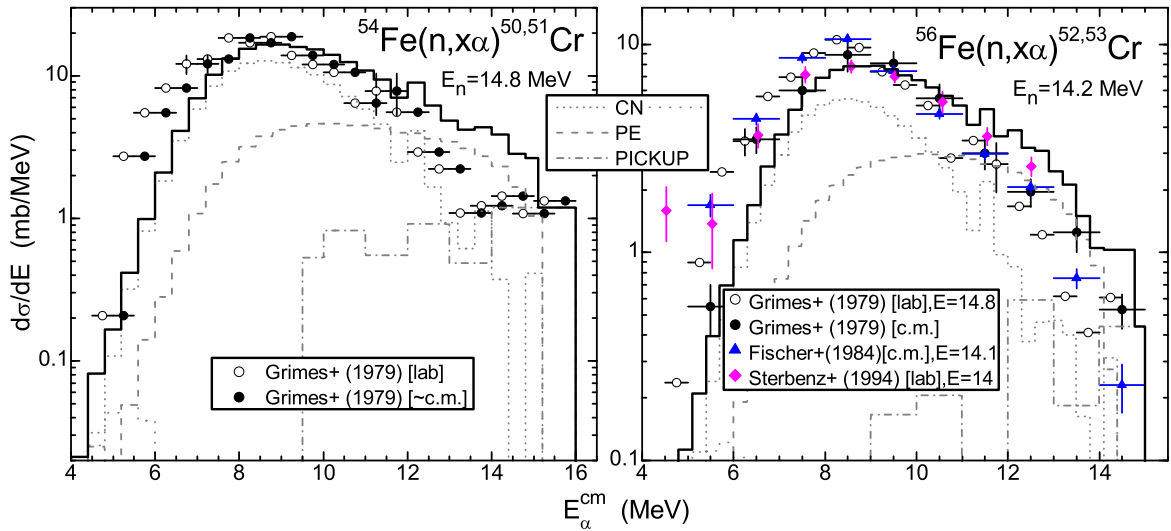}}
\caption{\label{Fig:Fe546nas} As Fig.~\ref{Fig:Co59naS} but for $^{54,56}$Fe target nuclei \cite{exforp,smg79p}, at incident energies of 14.8 and 14.2 MeV, respectively.}
\end{figure*}

\begin{figure*} 
\resizebox{1.8\columnwidth}{!}{\includegraphics{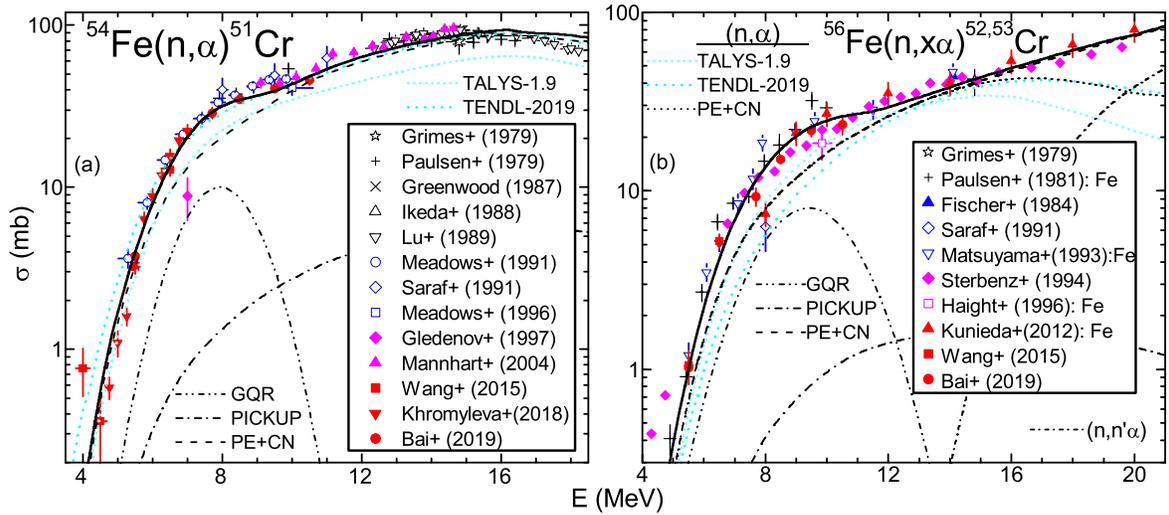}}
\caption{\label{Fig:Fe546na} As Fig.~\ref{Fig:Co59na} but for $^{54,56}$Fe target nuclei and additionally calculated $(n,\alpha)$ cross sections using TALYS-1.9 and its default parameters \cite{TALYSp} (short-dotted curves), and like--GQR components (dash-dot-dotted curves) added to DI+PE+CN (solid curves).}
\end{figure*}

At the same time, this spectrum analysis provides evidence for rather low value of the GDH pre-formation probability $\varphi$=0.04. 
An uncertainty band corresponding to half of it is assumed in the excitation function analysis for both total $\alpha$--emission and $(n,\alpha)$ reaction cross section (Fig.~\ref{Fig:Co59na}). 
The overlap of the related uncertainty band of calculated cross section and recent data is obvious.

On the other hand, Fig.~\ref{Fig:Co59na} proves also that there are no NLD effects below the incident energy of 10 MeV. 
The $\alpha$--particle OMP \cite{va14p} is confirmed, with the additional support of more data since earlier work \cite{va94p}.

{\it Like-GQR in addition to DI+PE+CN $\alpha$-emission.} -- 
The DI pickup contribution to $(n,\alpha)$ reaction on $^{54,56}$Fe corresponds to 36 excited states up to 5.943 MeV for the residual nucleus $^{51}$Cr, and 18 states up to 5.557 MeV for $^{53}$Cr \cite{SM2020p}.  
A first proof of the present approach was given by calculated angular distributions for the $\alpha$-energy bins from 6 to 14 MeV, compared to data of Fischer {\it et al.} \cite{rf84p} for $^{56}$Fe at 14.1 MeV (Fig. 23 of \cite{SM2020p}). 
It is of interest in the following to note the best agreement obtained for the medium-energy $\alpha$--emission, i.e. from 10 to 12 MeV.

The most significant feature of the comparison of measured and calculated $\alpha$--emission spectra (Fig.~\ref{Fig:Fe546nas}) is the DI pickup contribution on lowest states. 
Then, some spectra overestimation there is at the energies just below the DI maxima, at once with an underestimation of the low-energy side. 
However it could be related for $^{54}$Fe to a rough attempt to transform data from lab \cite{smg79p} to c.m. by an average shift of 0.5 MeV to higher energies, at variance to the precise one for $^{56}$Fe \cite{rf84p}. 
There are also known question marks at $\alpha$--particle energies below $\sim$6 MeV, pointed out by Fischer {\it et al.} \cite{rf84p}.
At the same time, the corresponding experimental total $\alpha$--emission cross sections are among the lowest data, at their incident energies, within the excitation functions (Fig.~\ref{Fig:Fe546na}). 

Nevertheless, the DI pickup contribution is improving the description of these excitation functions except the energy range below $\sim$12 MeV. 
The uncertainty bands related to number of low-lying levels, NLD, and PE parameters \cite{SM2020p} have already proved that none of their effects are in order at these energies. 
Moreover, the above-mentioned best agreement obtained for the angular distribution for $\alpha$-energy bin of 6-10 MeV, at 14.1 MeV incident energy on $^{56}$Fe, is additionally supporting both the CN approach and the involved  $\alpha$--particle OMP \cite{va14p}.

Following a former suggestion \cite{ma06p}, we have assumed that such reaction cross-section enhancement could be understood as decay from giant resonances populated via neutron capture. 
Although generally the decay of the GQR is observed with nucleon emission (\cite{
js81p} and Refs. therein), it has been shown (\cite{mf07p} and Refs. therein) that an appreciable non-statistical decay through $\alpha$--emission can occur. 
This assumption has been supported by the position of the extra yield, beyond DI+PE+CN component sum, just at the GQR energy $E_{GQR}$=$65A^{-1/3}$ MeV \cite{js81p}, i.e. 17.092 and 16.89 MeV for $^{55,57}$Fe excited nuclei, respectively. 
Therefore, we have obtained a fit of this extra yield by addition of Gaussian distributions at $E_{GQR}$, with FWHM widths of 2.35 and 3.54 MeV, and peak cross sections of 8 and 10 mb, respectively (Fig.~\ref{Fig:Fe546na}).

It is obvious that the widths of these Gaussian distributions are much lower than the systematic 'best' value $\Gamma$=$85A^{-2/3}$ MeV (Fig. 14 of Ref. \cite{js81p}). 
They are even lower than the $\Gamma$=$17.5A^{-1/3}$ MeV dependence that seems to describe better the inelastic hadron scattering data which were obtained taking into account also other GRs \cite{js81p}. 
Hence we would call these components only like--GQR contributions.
Further conclusions on the physics behind this empirical addition are not yet  evident, while more similar cases to be concerned may help. 
Nevertheless, a suitable account of the measured $(n,\alpha)$ reaction and $\alpha$-emission cross sections (Fig.~\ref{Fig:Fe546na}) is obtained by this additional like--GQR contribution which are larger than the DI pickup for incident energies $<$12 MeV.

{\it Conclusions.} -- 
A consistent set of model parameters, validated by analysis of various independent data, makes possible the assessment of an optical model potential \cite{va14p} also for nucleon-induced $\alpha$-emission for $A$$\sim$60 nuclei.
Particularly,  $\alpha$--emission from $^{55,57,58}$Fe and $^{60}$Co isotopes excited by $(n,\alpha)$ reaction, and $^{62,64,65,66}$Cu and $^{64,65,66,68}$Zn isotopes excited through both $(n,\alpha)$ and $(p,\alpha)$ reactions has been analyzed.
The advantage of rather recent data of low-lying states feeding is essential.
Further consideration of additional reaction channels leading to increase of the $\alpha$-emission cross sections beyond the statistical predictions has concerned the DI pickup and GQR similar features.\\  

This work was partly supported by Autoritatea Nationala pentru Cercetare Stiintifica (Project PN-19060102) and carried out within the framework of the EUROfusion Consortium and has received funding from the Euratom research and training programme 2014-2018 and 2019-2020 under grant agreement No 633053. The views and opinions expressed herein do not necessarily reflect those of the European Commission.

\newpage




\onecolumngrid
\begin{center}
{\bf \large Supplement to "Alternate Solution of the $\alpha$-Potential Mystery"}\\ 
\bigskip

V.~Avrigeanu\thanks{Electronic address: vlad.avrigeanu@nipne.ro} and 
M.~Avrigeanu\thanks{Electronic address: marilena.avrigeanu@nipne.ro}

{\it Horia Hulubei National Institute for Physics and Nuclear Engineering, P.O. Box MG-6, 077125 Bucharest-Magurele, Romania}
\end{center}

\twocolumngrid


Details of the consistent set of statistical-model input parameters in the main paper, and its validation by analyzing various independent data, are given in this supplement. 
The results obtained using an incident $\alpha$-particle optical model potential [Phys. Rev. C {\bf 90}, 044612 (2014)] within $A$$\sim$60 mass-number range, by taking the advantage of rather recent data of low-lying states feeding, are then discussed. 
Further consideration of an additional reaction channel leading to an increase of the $\alpha$-emission data beyond the statistical model predictions has also concerned the pickup direct interaction (DI). 
The assessment of DI cross sections has been subject to available information on  spectroscopic factors related to populated states, outgoing particle angular distributions, or at least differential cross--section maximum values. 


\section{INTRODUCTION} 

The $\alpha$-particle optical model potential (OMP) parameters are usually derived from analysis of either elastic-scattering, above the Coulomb barrier $B$, or $\alpha$-induced reaction data (e.g., \cite{pd02}) in terms of the statistical Hauser-Feshbach (HF) \cite{wh52} and pre-equilibrium emission (PE) \cite{eg92} models.  
Because there are also various assumptions and parameters of the HF+PE models in addition to the $\alpha$-nucleus potential, only the use of a consistent parameter set (e.g., \cite{eda80}) may provide a definite conclusion on the $\alpha$-particle OMP. 
Details of the consistent set of input parameters used within the main paper, and its validation by analyzing various independent data, are given in this supplement.
 
Moreover, the account at once of both absorption and emission of $\alpha$-particles in nuclear reactions \cite{tr13} may face the need for new physics in potentials to describe nuclear de-excitation (\cite{ma06} and Refs. therein). 
Therefore, it is of interest to see whether the incident $\alpha$-particle OMP \cite{va14} is able to describe also the $\alpha$-emission from excited nuclei \cite{ma06,va94} in reactions induced by low-energy protons \cite{va15} as well as fast neutrons \cite{va17}.
This aim could be particularly achieved within $A$$\sim$60 mass-number range, by taking the advantage of rather recent data of low-lying states feeding \cite{gz10a,gz10,yg14,zw15,tk18,hb19} also discussed in this supplement.  

Further consideration of an additional reaction channel \cite{tr13,va16a} but leading to an increase of the $\alpha$-emission data beyond the HF+PE predictions is also aimed in the following regarding the pickup direct interaction (DI) \cite{smq95,eg92,eg89,eg91}. 
An eventual giant quadrupole resonance (GQR) \cite{grs74,js81} decay through $\alpha$-particle emission (\cite{ns75,ew79,pg86,mf05,mf07} and Refs. therein) is discussed in the main paper. 

Nevertheless, before proceeding to the above-mentioned points, it should be mentioned that no empirical rescaling factors of the $\gamma$ and/or neutron widths are used in the present work as well as to establish this OMP \cite{va14,ma09,ma10} or prove it \cite{va16,va19}. 
First, its use provided a rather good agreement of the calculated and measured $\alpha$-induced reaction data to date \cite{va14,ma09,ma10,va15,va17,va16,va19} . 
Second, better results obtained also within large-scale nuclear-data evaluation \cite{ajk15} led to its adoption as the latest default option within the world-wide used code TALYS \cite{TALYS}. 
Consequently, conclusion on the suitable use of this $\alpha$-particle OMP also for the account of the $\alpha$-emission will be of a larger interest. 

Actually, a semi-microscopic double-folding model (DFM) real part and the dispersive contribution of a phenomenological energy-dependent imaginary-potential were firstly involved within an analysis of $\alpha$-particle elastic-scattering angular distributions above $B$ \cite{va14,ma09,ma03}. 
Subsequently, the phenomenological real potential \cite{va14} was established using the same data basis. 
Then, HF statistical model analysis of $\alpha$-induced reaction cross sections proved the particular energy dependence of the surface imaginary potential at incident energies below $B$ \cite{va14,ma09,ma10}. 

On the other hand, an earlier OMP \cite{va94} concerned only the $\alpha$-particle emission in neutron-induced reactions, with distinct predictions from potentials for incident $\alpha$ particles \cite{lmf66}. 
Thus, it was not an earlier version of the above-mentioned ones \cite{va14,ma09,ma03}, while the question on different OMPs for incident and emitted $\alpha$ particles \cite{ma06,va94,va15,va17} is still actual.

While detailed presentation of  model parameters was given in Refs. \cite{va14,va15,va16}, latest particular values of them are given in Sec.~\ref{SMcalc} of this supplement. 
The PE+HF results obtained using the OMPs of Refs. \cite{va14,va94} are then compared in Sec.~\ref{Res} with measured cross sections of low-energy proton-- and neutron--induced reactions leading to excited Fe, Co, Cu, and Zn isotopes, with particular emphasis of above-mentioned Refs. \cite{gz10a,gz10,yg14,zw15,tk18,hb19} in order to provide more details and support to the additional work within the main paper. 
The main points of the DI analysis within the distorted-wave Born approximation (DWBA) method and the code FRESCO \cite{FRESCO}, for calculation of pickup reaction cross sections, are also given in Sec.~\ref{DI}.

\section{Statistical model parameters} \label{SMcalc}

The following HF+PE model calculations were carried out within a local approach using an updated version of the computer code STAPRE-H95 \cite{ma95}, with $\sim$0.1-0.3 MeV equidistant binning for the excitation energy grid.
The DWBA method and a local version of the code DWUCK4 \cite{pdk84} were also used for calculation of the DI collective inelastic-scattering cross sections using the corresponding deformation parameters \cite{sr01,tk02} of the first 2$^+$ and 3$^-$ collective states.  
The calculated DI cross sections are then involved for the subsequent decrease of the total-reaction cross sections $\sigma_R$ that enter the HF calculations. 
Typical DI inelastic-scattering cross sections, e.g. for neutrons on $^{56}$Fe, grow up from $\sim$7\% to $\sim$10\% of $\sigma_R$ for incident energies from 2 to 7 MeV, and then decrease to $\sim$8\% at the energy of $\sim$25 MeV.

A consistent set of (i) back-shifted Fermi gas (BSFG) \cite{hv88} nuclear level density (NLD) parameters, (ii) nucleon and (iii) $\gamma$-ray transmission coefficients was used also within the present analysis of the fast-neutron induced reactions on isotopes of Fe, Co, Cu, and Zn. 
These parameters were established or validated using independently measured data as 
low-lying levels \cite{ensdf} and $s$-wave nucleon-resonance spacings, $(p,n)$ reaction cross sections \cite{exfor}, radiative strength functions (RSF) \cite{OCL}, average $s$-wave radiation widths \cite{ripl3}, and $(p,\gamma)$ reaction cross sections \cite{exfor}, respectively. 
The same OMP and level density parameters have been used in the framework of the DI, PE, and HF models. 
The details in addition to the ones given formerly \cite{va14,va15,va17,va16,va19} as well as particular parameter values are mentioned below in order to provide the reader with all main details and assumptions of the present analysis.

The reaction cross sections calculated within this work are also compared with the TALYS code results corresponding to its default options as well as the content of the evaluated data library TENDL-2019 \cite{TENDL}, for an overall excitation function survey. 

\subsection{Nuclear level densities} \label{NLD}

The BSFG parameters used to obtain the present HF results are given in Table~\ref{densp}. They follow the low-lying level numbers and corresponding excitation energies used in the HF calculations (the 2nd and 3rd columns) \protect\cite{ensdf} as well as those fitted at once with the available nucleon-resonance data \cite{hv88,ripl3,ll08,ripl2} to obtain these parameters. 
The level-density parameter $a$ and ground state (g.s.) shift $\Delta$ were generally obtained with a spin cutoff factor corresponding to a variable moment of inertia $I$, between half of the rigid-body value $I_r$ at  g.s., 0.75$I_r$ at the separation energy $S$, and the full $I_r$ value at the excitation energy of 15 MeV, with a reduced radius $r_0$=1.25 fm \cite{va02}.
 
\begingroup
\squeezetable
\begin{table*} 
\caption{\label{densp} Low-lying levels number $N_d$ up to excitation energy $E^*_d$ \protect\cite{ensdf} used in HF calculations, $N_d(E^*_d)$ and $s$-wave nucleon-resonance spacings$^a$ $D_0^{exp}$ (with uncertainty in units of the last digit, in parentheses) in one or more (separated by slash) energy ranges $\Delta$$E$ \cite{ripl1b} above separation energy $S$, for the target--nucleus g.s. spin $I_0$, fitted to obtain LD parameter {\it a} and g.s. shift $\Delta$ (for a spin cutoff factor related to a variable moment of inertia \cite{va02} between half and 75\% of the rigid-body value, from g.s. to $S$, and reduced radius $r_0$=1.25 fm) with uncertainty related firstly to those of fitted $D_0^{\it exp}$ and, in addition, to those of fitted $N_d$ (in second pair of brackets).} 
\begin{ruledtabular}
\begin{tabular}{cccccccccc} 
Nucleus   &$N_d$&$E^*_d$& \multicolumn{5}{c}
                  {Fitted level and resonance data}& $a$ & $\Delta$\hspace*{3mm}\\
\cline{4-8}
           &  &     &$N_d$&$E^*_d$&$S+\frac{\Delta E}{2}$&
                                     $I_0$&$D_0^{\it exp}$ \\ 
           &  &(MeV)&   & (MeV)& (MeV)&  &(keV)&(MeV$^{-1}$) & (MeV) \\ 
\noalign{\smallskip}\hline\noalign{\smallskip}
$^{49}$V &25&2.408&25   &2.408& 9.559& 0 & 10.6(10)$^b$ & 5.35 &-1.80 \\
$^{50}$V &43&2.534&43   &2.534&      &   &              & 5.95 &-1.75 \\
$^{51}$V &55&4.053&55   &4.053&11.071 / 10.646& 6 / 0 &2.0(1) / 7.9(6)$^b$& 5.56 &-0.85 \\
$^{52}$V &32&2.591&32   &2.591& 7.361&7/2&  4.0(6)      & 6.33 &-1.13 \\
$^{53}$V &20&2.772&24   &2.967&      &   &              & 5.75 &-0.90 \\
$^{54}$V &17&1.540&17   &1.540&      &   &              & 6.20 &-2.20 \\
$^{49}$Cr&13&2.613&12   &2.578&      &   &              & 5.50 &-0.70 \\
$^{50}$Cr&28&4.367&28   &4.367&      &   &              & 5.60 & 0.31 \\
$^{51}$Cr&41&3.376&82(2)&4.214& 9.655& 0 &12.5(12),13.1(12)$^c$&5.43(10)(8)&-1.30(12)(6)\\
         &  &     &     &     & 9.561& 0 &13.3(13)$^b$ \\
$^{52}$Cr&36&5.139&36   &5.139&      &   &              & 5.30       & 0.47\\
$^{53}$Cr&28&3.435&25(2)&3.262& 8.432 / 8.654& 0 &32.0(35) / 30.1(35)$^c$& 5.55(12)(6)&-0.80(10)(-5)\\
$^{54}$Cr&38&4.689&38   &4.689& 9.817 / 10.001&3/2&  6.7(6) / 5.81(59)$^c$&5.65(10)&0.24(9)\\
$^{55}$Cr&32&3.200&32   &3.200& 6.396& 0 &  50(8)       &6.47(24)    &-0.50(15)\\
         &  &     &     &     &6.696 / 6.663& 0 &  54.4(82)$^b$ / 52.6(77)$^c$ \\
$^{56}$Cr&29&4.349&29   &4.349&      &   &              & 5.85       & 0.32 \\
$^{52}$Mn&15&1.956&15   &1.956&      &   &              & 6.10       &-1.20 \\
$^{53}$Mn&33&3.466&42   &3.728&      &   &              & 5.60       &-0.85 \\
$^{54}$Mn&36&2.355&36(3)&2.355&      &   &              & 5.7(2)     &-2.02(9) \\
$^{55}$Mn&58&3.383&58   &3.383&10.497& 0 &  7.1(7)$^b$  & 5.53       &-1.69 \\
$^{56}$Mn&49&2.118&49(2)&2.118& 7.374&5/2&2.3(4)&6.20(42/-3)(37/0)&-2.22(32/-3)(24/4)\\
$^{57}$Mn&23&2.341&33   &2.758&      &   &              & 6.0(2)       &-1.32 \\
$^{58}$Mn&27&1.470&27   &1.470&      &   &              & 6.40       &-2.25 \\
$^{52}$Fe& 9&4.456& 5   &3.585&      &   &              & 5.30       & 1.15 \\
$^{53}$Fe&24&3.176&24(2)&3.176&      &   &              & 5.47(20)   &-0.90(8)\\
$^{54}$Fe&26&4.782&26(2)&4.782&      &   &              & 5.65(20)   & 0.76(8)\\
$^{55}$Fe&31&3.457&31(2)&3.457& 9.548& 0 &18.0(24)$^b$,20.5(14)&5.53(14)(10)&-0.90(12)(2)\\
$^{56}$Fe&60&5.038&49(2)&4.802&      &   &              & 6.0(2)     & 0.36(17)(12)\\
$^{57}$Fe&30&2.697&30(2)&2.697&8.074 / 8.096& 0 & 19.2(19)$^b$ / 25.4(19)& 6.00(11)(6)&-1.28(8)(3) \\
$^{58}$Fe&60&4.720&60   &4.720&10.14 &1/2& 7.05(70)     & 5.90       &-0.90 \\
$^{59}$Fe&32&2.570&36   &2.856&6.756 / 6.696& 0 &21.6(26) / 25.4(49)$^b$& 6.16       &-1.03 \\
$^{60}$Fe&34&4.053&34   &4.053&      &   &              & 6.25       & 0.11 \\
$^{61}$Fe&16&2.143&16   &2.143&      &   &              & 7.00       &-0.72 \\
$^{55}$Co&29&3.980&23   &3.775&      &   &              & 5.40       &-0.33 \\
$^{56}$Co&27&2.789&27   &2.789&      &   &              & 6.40       &-0.88 \\
$^{57}$Co&44&3.553&44   &3.553&      &   &              & 5.80       &-0.97 \\
$^{58}$Co&43&1.979&43(2)&1.979&      &   &              & 6.6(2)     &-2.11(10)\\
$^{59}$Co&60&3.492&61   &3.497&10.217& 0 & 4.3(4)$^b$   & 6.40       &-0.90 \\
$^{60}$Co&60&2.423&64   &2.489& 7.542&7/2& 1.45(15)     &6.99(14)    &-1.57(10)\\
$^{61}$Co&60&3.417&70(2)&3.575&      &   &              & 6.45(35)   &-0.96(24)\\
$^{62}$Co&17&1.359&16   &1.271&      &   &              & 7.40       &-1.48 \\
$^{63}$Co&11&2.163& 8   &1.889&      &   &              & 7.30       &-0.30 \\
$^{58}$Ni&34&4.574&34   &4.574&      &   &              & 5.90       & 0.40 \\ 
$^{59}$Ni&44&3.381&58   &3.686&9.405 / 9.324& 0 &13.4(9) / 12.5(9)$^b$& 5.85 &-1.15 \\
$^{60}$Ni&51&4.613&51   &4.613&      &   &  2.0(7)      & 6.00       & 0.06 \\ 
$^{61}$Ni&36&2.913&36   &2.913& 8.045& 0 &13.8(9),13.9(15)$^b$&6.57(9)&-0.90(7)\\ 
$^{62}$Ni&46&4.455&46   &4.455&10.631&3/2&  2.10(15)    & 6.36       & 0.27 \\ 
$^{63}$Ni&19&2.353&19   &2.353& 7.117 / 7.238& 0 &16(3) / 15(2)$^b$ & 7.30 &-0.52 \\ 
$^{64}$Ni&20&3.849&20   &3.849&      &   &              & 6.90       & 0.75 \\ 
$^{65}$Ni&20&2.520&20   &2.520& 6.398& 0 &  23.6(30)    & 7.80       &-0.20 \\ 
$^{66}$Ni&18&3.782&18   &3.782&      &   &              & 7.50       & 1.12 \\ 
$^{59}$Cu&38&3.758&38   &3.758&      &   &              & 6.30       &-0.23 \\
$^{60}$Cu&24&1.505&24   &1.505&      &   &              & 7.00       &-1.75 \\ 
$^{61}$Cu&36&3.042&38   &3.092&      &   &              & 6.75       &-0.64 \\
$^{62}$Cu&48&1.682&53   &1.775&      &   &              & 7.20       &-2.04 \\ 
$^{63}$Cu&48&3.043&48   &3.043& 9.026& 0 &  5.9(7)$^b$  & 6.90       &-0.83 \\ 
$^{64}$Cu&60&2.115&93   &2.534& 7.993&3/2&0.95(9)$^d$ [0.70(9)]      &7.46&-1.61\\ 
$^{65}$Cu&40&3.132&48   &3.278&      &   &              & 7.85       &-0.10 \\ 
$^{66}$Cu&20&1.344&20   &1.344& 7.116&3/2&  1.30(11)    & 7.88       &-1.40 \\ 
$^{67}$Cu&12&2.841&18   &3.123&      &   &              & 8.20       & 0.62 \\ 
$^{61}$Zn&15&1.730&15   &1.730&      &   &              & 6.70       &-1.25 \\ 
$^{62}$Zn&23&3.730&23   &3.730&      &   &              & 6.50       & 0.32 \\ 
$^{63}$Zn&32&2.403&32   &2.403&      &   &              & 7.50       &-0.79 \\ 
$^{64}$Zn&34&3.552&68   &4.159&11.862&3/2&              & 7.20       & 0.16 \\ 
$^{65}$Zn&33&2.138&33   &2.138& 8.018& 0 &  2.3(3)      & 8.29       &-0.77 \\
$^{66}$Zn&39&3.825&39   &3.825&      &   &              & 7.70       & 0.55 \\
$^{67}$Zn&40&2.175&40   &2.175& 7.278& 0 & 4.62(55)     & 8.11       &-0.97 \\ 
$^{68}$Zn&51&3.943&51   &3.943&10.291&5/2& 0.37(2)      & 8.05       & 0.59 \\ 
\end{tabular}	 
\end{ruledtabular}
\begin{flushleft}
$^a$Reference \cite{ripl3} if not otherwise mentioned.
\hspace*{0.5in} $^b$Reference \cite{hv88}
\hspace*{0.5in} $^c$Reference \cite{ll08}
\hspace*{0.5in} $^d$Reference \cite{ripl2}\\
\end{flushleft}
\end{table*}
\endgroup

The fit of the error-bar limits of $D_0^{\it exp}$ data has also been used to provide limits of the fitted $a$-parameters. 
Moreover, these limits are used within HF calculations to illustrate the NLD effects on the calculated cross-section uncertainty bands (Sec.~\ref{Res}). 

On the other hand, the smooth-curve method \cite{chj77} was applied for nuclei without resonance data.
Thus, average $a$-values of the neighboring nuclei having these data hsve been used to obtain only the $\Delta$ values by fit of the low-lying discrete levels. 
The uncertainties of these averaged $a$-values, following the spread of the fitted $a$ parameters, are also given in Table~\ref{densp}.
These uncertainties are obviously larger than those of the $a$-values obtained by fit of $D_0^{\it exp}$. 
Their use in HF calculations leads to increased calculated cross-section uncertainty bands.
Moreover, eventual assumption of an additional uncertainty for the fitted $N_d$ has led to increased NLD parameter uncertainties as given in a second pair of brackets in Table~\ref{densp}.

\subsection{Nucleon optical model potentials} \label{OMP}

\subsubsection{Neutron optical model potentials} \label{OMPn}

{\it Fe isotopes'} first option for neutron OMP was obviously the optical potential of Koning and Delaroche \cite{KD03}. 
However, we paid the due attention to the authors' remark that their global potential does not reproduce the minimum around the neutron energy of 1--2 MeV for the neutron total cross sections of the $A$$\sim$60 nuclei. 
Following also their comment on the constant geometry parameters which may be responsible for this aspect, we applied the SPRT method \cite{jpd76} for determination of these parameters at energies below $\sim$20 MeV, through analysis of the $s$- and $p$-wave neutron strength functions, the potential scattering radius $R'$ and the energy dependence of neutron total cross section $\sigma_T (E)$. 

\begin{figure*} 
\resizebox{2.0\columnwidth}{!}{\includegraphics{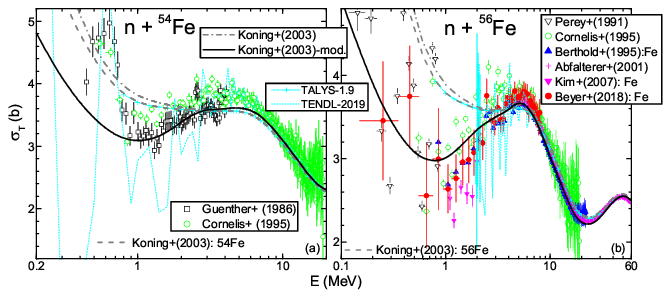}}
\caption{\label{Fig:Fe546nT} (Color online) Comparison of neutron total cross sections for $^{54,56}$Fe measured \cite{exfor} and calculated using either the global (dash-dotted curves) or local (dashed curves) OMP parameters sets of Koning and Delaroche \cite{KD03}, and the energy-dependent changes of either (a) global or (b) local geometry parameters given in Table~\ref{tab:nres} (solid curves). Broad energy--averages over 50, 100, and particularly \cite{rb18} 200 keV of several measured data sets were used for comparison with the OMP results.} 
\end{figure*}

The RIPL-3 recommendations \cite{ripl3} for the neutron resonance data and the available measured $\sigma_T$ data \cite{exfor} (Fig.~\ref{Fig:Fe546nT}) have been used in this respect.
One may note that more accurate data have become available since the development of Koning--Delaroche OMP. 
Moreover, it has been shown that even recent dispersive-coupled-channels OM results describing well the total cross sections above 4 MeV, are still rather similar to \cite{KD03} predictions at lower energies \cite{rb18}. 
The model overestimation around 1 MeV is up to $\sim$20 \%. 

\begin{table*} 
\caption{\label{tab:nres}Comparison of experimental \cite{ripl3} and calculated neutron scattering parameters of $^{54,56}$Fe isotopes at neutron energies of 250 and $\sim$400 keV \cite{ripl1b}, respectively, and (bottom) the changes of the global ($^{54}$Fe) and local ($^{56}$Fe) parameters \cite{KD03} (with the use of \cite{KD03} notations, the energies are in MeV and geometry parameters in fm) which provide the best SPRT results.}
\begin{tabular}{lccc|ccc} 
\hline
\hspace*{0.25in} $\backslash$ Target 
       & \multicolumn{3}{c}{$^{54}$Fe}
       & \multicolumn{3}{c}{$^{56}$Fe} \\
\cline{2-4}   \cline{5-7} 
OMP & S$_0$*10$^4$ & S$_1$*10$^4$ & R' & 
      S$_0$*10$^4$ & S$_1$*10$^4$ & R' \\
\hline
Exp. \cite{ripl3}                & 6.6(10) & 0.42(8) &     & 2.3(6) & 0.41(6) & \\
global \cite{KD03} &  2.74   &  0.73   & 5.2 &  2.24  &   0.7   & 4.7 \\
local \cite{KD03} &  2.68   &  0.82   & 4.9 &  2.28  &   0.7   & 4.5 \\
\cite{KD03} modified &  2.25   &  0.45   & 4.1 &  1.72  &   0.46  & 3.5 \\
\cline{2-4}   \cline{5-7} 
       & \multicolumn{3}{c}{$r_V$$=$1.2766-0.04$E$, $E$$<$2}
       & \multicolumn{3}{|c}{$r_D$$=$1.282+0.014$E$, $E$$<$5} \\
			 & \multicolumn{3}{c}{ }
			 & \multicolumn{3}{|c}{\hspace*{0.2in}1.502-0.03$E$, $E$$<$10} \\  
			 & \multicolumn{3}{c}{ }
			 & \multicolumn{3}{|c}{\hspace*{0.33in}1.182+0.002$E$, $E$$>$10} \\  
       & \multicolumn{3}{c}{$a_V$$=$0.2198+0.15$E$, $E$$<$3}
       & \multicolumn{3}{|c}{$a_V$$=$0.333+0.066$E$, $E$$<$5} \\
\hline
\end{tabular}
\end{table*}
\begin{figure*} 
\resizebox{2.0\columnwidth}{!}{\includegraphics{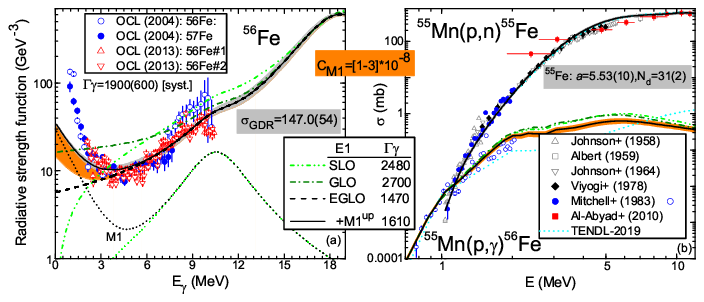}}
\caption{\label{Fig:RSFMn55pn} (Color online) Comparison of (a) measured \cite{OCL} and sum of calculated $M$1-radiation SLO model (dotted curve) with E1-radiation SLO (dash-dot-dotted curve), GLO (dash-dotted curve), and EGLO models (dashed curve) RSFs, as well as the sum (solid curve) of the $M$1 component including the upbend to zero energy (short dotted curve) and EGLO, for $^{56}$Fe nucleus, and 
(b) cross sections for $(p,\gamma)$ and $(p,n)$ reactions on $^{55}$Mn, measured \cite{exfor}, evaluated \cite{TENDL}, and calculated using the above-mentioned E1 radiation RSF models and curves for the former;
uncertainties correspond to (a) E1-radiation EGLO/GDR parameters (light-gray band) and, in addition, for the M1-radiation upbend (orange band), and 
(b) the above-mentioned RSF uncertainties (orange band), for $(p,\gamma)$ reaction, and $^{55}$Fe levels and NLD parameter uncertainties (light-gray band) for $(p,n)$ reaction.
The $s$-wave average radiation widths $\Gamma_{\gamma}$ (in meV) are either deduced from systematics \cite{ripl3} or correspond to M1 and each of above-mentioned E1 models.} 
\end{figure*}

Actually, it has been supported the well known behavior for near--magic nuclei of the iron group that the observed reduction in the averaged $\sigma_T (E)$ requires $l$-dependent OMPs \cite{tk97}.
We found however that it is possible to describe somehow the $\sigma_T (E)$ minimum around 1 MeV by adoption of the energy--dependent geometry parameters of either the  global ($^{54}$Fe) or local ($^{56}$Fe)  Koning--Delaroche potentials, at lower energies, given in Table~\ref{tab:nres}. 
This point will be of a particular importance for the correct account of the neutron evaporation in competition with the charged--particle emission.

{\it Cu isotopes} have shown a rather likewise case, however with a $\sigma_T (E)$ minimum around 2 MeV better approximated by Koning--Delaroche OMP. A similar analysis as above  is given in \cite{ma08}, and their similarly modified local potentials for $^{63,65}$Cu \cite{KD03} have used in the present work as well.

{\it Zn isotopes} proves an even better account of the $\sigma_T (E)$ minimum above 2 MeV but a minor underestimation of $\leq$3\% for the maximum around 6 MeV. 
Both points have been improved through the SPRT method, by using Koning--Delaroche local parameter set for $^{74}$Ge \cite{KD03} with a minor change of the real potential radius $r_V$$=$1.214-0.001$E$, for $E$$<$6 MeV.

{\it $^{59}$Co target nucleus} made the object of a detailed study of neutron-induced reactions \cite{vs04} that was completed at nearly the same time with Koning--Delaroche OMP. However, a consistent input parameter set was also used so that an earlier neutron OMP \cite{rdl88} was adopted with changes according to the SPRT method as well. Because of the completeness of that study, only an additional analysis of the $\alpha$-particle emission has been carried on in the present work, with no further change of the adopted nucleon OMPs \cite{vs04}.

\subsubsection{Proton optical model potentials} \label{OMPp}

The optical potential of Koning and Delaroche \cite{KD03} was considered also for the calculation of proton transmission coefficients on isotopes of Mn, Fe, Ni, and Cu. 
A former trial of this potential concerned the proton reaction cross sections $\sigma_R$ \cite{rfc96} for the stable isotopes of the elements from Mn to Zn, for the lower energies so important in statistical emission from excited nuclei. 
A comparison of these data and results of either the local or global proton OMP \cite{KD03} was shown in Fig. 2 of Ref. \cite{ma08}. 
A good agreement was found apart from the isotopes of Fe and in particular Ni, for which there is an overestimation of the data by $\geq$10\%. 
The following additional remarks concern the isotopes involved in this work.

{\it Fe and Ni isotopes} could benefit from the above-mentioned work \cite{ma08} where the agreement with the corresponding $\sigma_R$ data was achieved using energy-dependent real potential diffusivity. 
Thus, the constant value $a_V$=0.663 fm \cite{KD03} was replaced by the energy--dependent form 
$a_V$=0.563+0.002$E$ up to 50 MeV, for the target nucleus $^{56}$Fe, and $a_V$=0.463+0.01$E$ up to 20 MeV for $^{58}$Ni. 
A final validation of the additional energy--dependent $a_V$ was obtained for Ni isotopes of interest for this work, by analysis of the available $(p,\gamma)$ and $(p,n)$ reaction data up to $E_p\sim$12 MeV (Fig. 4 of \cite{ma08}), the other HF parameters being the same as in the rest of the previous as well as present work.

Consequently the same modified OMP has been used in this work for Fe isotopes. 
On the other hand, the enlarged incident-energy range for $(p,n)$ reaction on $^{61,64}$Ni in the present work (Sec.~\ref{Res}) led only to a minor change of the local parameter sets for $^{60,64}$Ni \cite{KD03}. 
Thus, the energy-dependent $a_V$=0.463+0.004$E$ up to 50 MeV has been used within the analysis of the above-mentioned reaction. 
Discussion of the corresponding calculated cross sections is given in the following section.

{\it Cu isotopes} made the object of a similar analysis in the meantime \cite{va16}, to validate the proton OMP within an analysis of quite accurate $(\alpha,x)$ data at lowest energies, where $x$ stands for $\gamma$, $n$, and $p$. 
Following the results shown in Fig. 1 of Ref. \cite{va16}, the same local OMP \cite{ss83} has been used also in this work. 

{\it Mn isotopes} have been additionally concerned in the present work due to the scarce related $\sigma_R$ data \cite{rfc96} and the well-known anomalies of proton OMP within these energies and mass range \cite{sk79}.
Thus, an analysis of the $(p,\gamma)$ and  $(p,n)$ reaction cross sections was carried out for $^{55}$Mn target nucleus and incident energies below $\sim$10 MeV,  with the results shown in Fig.~\ref{Fig:RSFMn55pn}(b).  
The corresponding HF calculations were obviously carried out using the same input parameters as in the rest of this work and the similar ones for Cu \cite{va16} and Zn \cite{va15} isotopes. 
While more details are given below for the analysis of the RSF also shown in Fig.~\ref{Fig:RSFMn55pn}(a), a rather good agreement has been obtained using the proton OMP global parameters  \cite{KD03}.
This result is particularly obvious for the $(p,\gamma)$ reaction below the $(p,n)$ reaction effective threshold, where its cross section is closest to $\sigma_R$. 

One may also note that the NLD parameter uncertainty related to the limits of the fitted $D_0^{\it exp}$ and low-lying levels of $^{55}$Fe (Table~\ref{densp}) provides 
no significant effect on the calculated $(p,n)$ cross sections. Therefore the OMP global parameters \cite{KD03} have been used for protons on Mn isotopes in this work.

\subsection{$\gamma$-ray strength functions} \label{RSF}

\begin{figure} 
\resizebox{1.0\columnwidth}{!}{\includegraphics{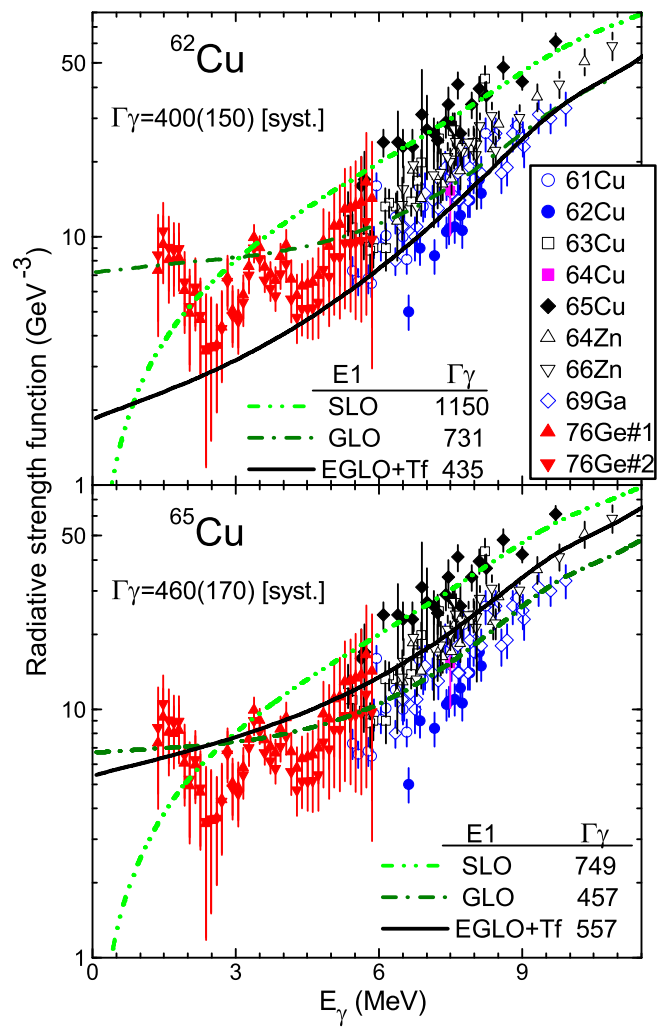}}
\caption{\label{Fig:RSF_Cu6265} (Color online) Comparison of the sum of calculated $\gamma$-ray strength functions of the E1 and M1 radiations for $^{62,65}$Cu nuclei,  using to the SLO (dash-dot-dotted curves), GLO (dash-dotted curves), and EGLO (solid curves) models for E1 radiations, and SLO model for M1 radiations; 
$s$-wave average radiation widths $\Gamma_{\gamma}$ (in meV) are either deduced from systematics \cite{ripl3} or correspond to $M1$ and each of above-mentioned E1 models.  
There are also shown the measured dipole $\gamma$-ray strength functions for $^{61,62}$Cu \cite{kn82}, $^{63}$Cu \cite{kn80}, $^{64}$Cu \cite{ripl2}, $^{65}$Cu \cite{be79}, $^{64,66}$Zn \cite{be80}, $^{69}$Ga \cite{kn83}, and $^{76}$Ge nuclei \cite{as14}.
.}
\end{figure}

Measured RSF and average $s$-wave radiation widths $\Gamma_{\gamma}$ \cite{ripl3} data have already been largely used for the suitable account of $\gamma$-ray transmission coefficients for compound nuclei (CN) with $A$$\geq$60 \cite{va16}. 
Nevertheless, the isotopes $^{56,57}$Fe are most significant for the present work, because of an unexpected RSF upbend to zero energy was firstly discovered for them \cite{av04}. 
However, despite more detailed studies of additionally related data (\cite{acl17} and Refs. therein), there is not yet an unique parametrization for these nuclei. 
Therefore, firstly, we have adopted the recently-compiled \cite{RIPL3gamma,tk20} giant dipole resonance (GDR) parameters within the former Lorentzian (SLO) \cite{pa62}, generalized Lorentzian (GLO) \cite{jk90}, and enhanced generalized Lorentzian (EGLO) \cite{jk93} models for the electric-dipole RSF. 
The constant nuclear temperature $T_f$=1.2 MeV of the final states \cite{acl10} was particularly assumed within the EGLO model.

Then, the SLO model has been used for the M1 radiation, with the global \cite{ripl3} GDR energy and width, i.e. $E_0$=41/$A^{1/3}$ MeV and $\Gamma_0$=4 MeV.
The related peak cross section $\sigma_0$=2 mb has been assumed at once with the above-mentioned $T_f$ value in order to describe the RSF data \cite{OCL} around the $M1$ peak as shown in Fig.~\ref{Fig:RSFMn55pn}(a). 
Moreover, an additional M1 upbend to zero energy has been described by the function $f_{up}(E_{\gamma})$=$C$exp$(-{\eta}E_{\gamma})$, with the parameter value $\eta$=0.8MeV$^{-1}$ and limits $C$=(1--3)$\times$10$^{-8}$ MeV$^{-3}$ \cite{sg18} [orange band in Fig.~\ref{Fig:RSFMn55pn}(a)]. 
The RSF data are notably underestimated below $\sim$2 MeV (see also \cite{sg18}) while a better trend is provided in comparison with the sum of the SLO for $M1$ radiation and either SLO or GLO models for the $E1$ radiation.

There is shown in the same figure the uncertainty propagation of the electric-dipole GDR parameter $\sigma_0$=(147$\pm$5.4) mb \cite{RIPL3gamma} and M1 upbend parameter $C$ on the RSF energy dependence. 
While the former is important mainly above the nucleon binding energy but yet within the RSF-data errors, the latter is obviously essential below $E_{\gamma}$$\sim$5 MeV.
Nevertheless the final propagation of the RSF uncertainty on the calculated $(p,\gamma)$ reaction cross sections corresponds to variations well within the data spreading as well as the effects due to various electric-dipole RSF models.

On the other hand, a comparison of the systematic estimation \cite{ripl3,acl17} and calculated $\Gamma_{\gamma}$ values corresponding to the three electric-dipole RSF models, as well as to addition of the M1 upbend to the EGLO model, is also shown in Fig.~\ref{Fig:RSFMn55pn}(a). 
Obviously, it provides an increased support to the latest E1 model.

Therefore, the present results of RSF data analysis for $^{56}$Fe and related uncertainty propagation on calculated reaction cross sections strengthen the previous RSF approach for $A$$\geq$60 (Fig. 2 of Ref. \cite{va16}). 

Use of the above-mentioned electric-dipole RSF models is also discussed in Sec.~\ref{Res} for proton-induced reactions on $^{61,64}$Ni while the corresponding RSFs are shown in Fig.~\ref{Fig:RSF_Cu6265}. 
There is an obvious difference between the measured RSF data that are smaller for $^{62}$Cu \cite{kn82} than for $^{65}$Cu \cite{be79}.
We have described it within the EGLO model using different values of the constant nuclear temperature $T_f$.
Thus, 0.7 MeV has been considered for $^{62}$Cu, as for Zn isotopes \cite{va16}, but 1.2 MeV for $^{65}$Cu.
However, the agreement of the measured and our calculated RSF values has been considered at once with that for $\Gamma_{\gamma}$ values that are also shown in Fig.~\ref{Fig:RSF_Cu6265}. 
Actually, it is of interest to look for the effects on calculated cross sections in Sec.~\ref{Res} of the possible RSF uncertainty between the results of EGLO and GLO models, for $^{62}$Cu, and EGLO and SLO models, for $^{65}$Cu.

\subsection{Pre-equilibrium emission modeling} \label{PE}

The Geometry-Dependent Hybrid (GDH) model \cite{mb83}, generalized through inclusion of the angular-momentum and parity conservation \cite{ma90} and knockout $\alpha$-particle emission based on a pre-formation probability $\varphi$ \cite{eg92}, has been involved within STAPRE-H95 code to provide the PE contribution to the results of the present work. 
It includes also a revised version of the advanced particle-hole level densities (PLD) \cite{ma98,ah98} using the linear energy dependence of the single-particle level density \cite{mh89}. The particular energy dependence of the PE contribution within this approach, that is discussed at large in Sec. III.B.5 of Ref. \cite{pr05} for neutron-induced reactions on Mo isotopes, is thus fully appropriate also to this work. 
The central-well Fermi energy value $F$=40 MeV has been used, while the local-density Fermi energies corresponding to various partial waves (e.g., Fig. 4 of Ref. \cite{pr05}) were provided within the local density approximation by the same OMP parameters given in Sec.~\ref{OMP}. 

\section{$PE+HF$ Results and Discussion} \label{Res}

The PE+HF results corresponding to use of the consistent parameter set mentioned above (Sec.~\ref{SMcalc}) are compared in the following with measured cross sections of low-energy proton- and neutron-induced reactions leading to excited Fe, Co, Cu, and Zn isotopes. 
A particular attention will be payed to the recent data, e.g., \cite{gz10a,gz10,yg14,zw15,tk18,hb19}. 
The aim is to ascertain either the account of the $\alpha$-particle emission by the $\alpha$-particle OMP \cite{va14} or eventual questions that may still need further consideration.
Nevertheless, the suitable description of all available data for competitive reaction channels is firstly concerned, in order to avoid less accurate parameter-error compensation effects.

\subsection{Fe isotopes de--excitation} \label{Fe}

\begin{figure} 
\resizebox{1.0\columnwidth}{!}{\includegraphics{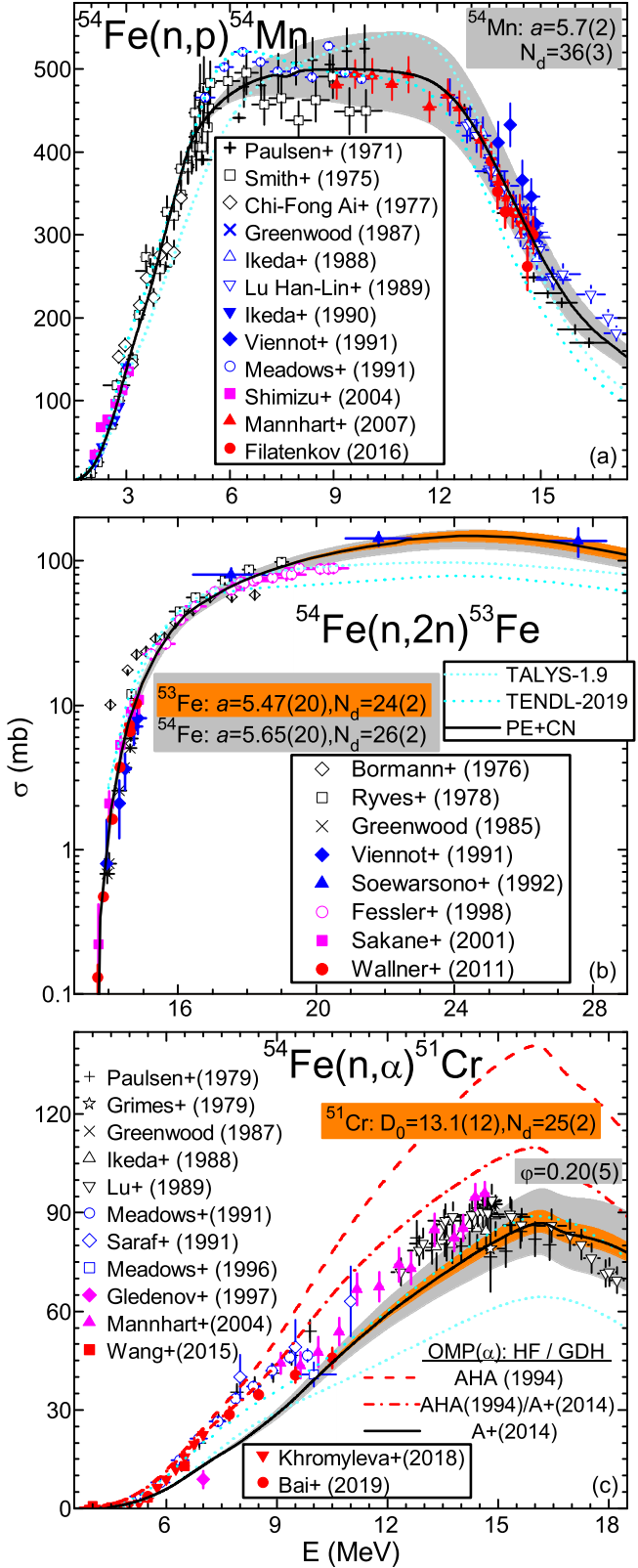}}
\caption{\label{Fig:Fe54nx} (Color online) Comparison of measured \cite{exfor}, evaluated \cite{TENDL} (short-dashed curves), and calculated TALYS-1.9 (short-dotted curves), and this work (solid curves) cross sections of neutron--induced reactions on $^{54}$Fe, with (c)  alternate use of $\alpha$-particle OMP \cite{va94} in HF (dash-dotted curve) and both HF+GDH (dashed curve) calculations; uncertainty bands correspond to error bars of $N_d$, and either LD parameter $a$ or $D_0^{\it exp}$ (Table~\ref{densp}) of residual nuclei (a) $^{54}$Mn, (b) $^{53}$Fe (orange band) and, in addition, $^{54}$Fe (gray band), and (c) $^{51}$Cr (orange band), as well as for GDH $\alpha$-particle pre-formation probability $\varphi$=0.20$\pm$0.05 (gray band).}
\end{figure}

The Fe isotopes and particularly the more abundant $^{54,56}$Fe are among the best studied nuclei also for the neutron-induced reactions (e.g., \cite{mh18} and Refs. therein). 
There are, however, still open question motivating quite useful recent studies as, e.g., \cite{rb18,zw15,tk18,hb19}. 
Moreover, there are interesting issues related to the neutron magic number $N$=28 of $^{54}$Fe as the striking difference between the $(n,p)$ excitation functions for the target nuclei $^{54,56}$Fe. 
On the other hand, there are recently measured essential cross sections for populations of the first three discrete levels of the even-even residual nucleus $^{54}$Cr by $(n,\alpha)$ reaction on $^{57}$Fe \cite{yg14}.
Their suitable description by model calculations is therefore a key test of the model assumptions and performance.

\subsubsection{$^{55}$Fe de--excitation} \label{54Fe}

{\it The $(n,p)$ reaction} large cross sections for this semi-magic target nucleus provides from the beginning an useful check of the proton optical potential. 
Thus, no NLD effects, including an uncertainty of the low-lying levels, can be seen in Fig.~\ref{Fig:Fe54nx}(a) until an incident energy of $\sim$5 MeV. 

The accuracies assumed for the number of low-lying levels and level density parameter $a$ of the residual nucleus $^{54}$Mn are then leading to an uncertainty of up to $\sim$10\% that may explain also the variance of the TALYS and TENDL-2019 results.
The last comment concerns firstly the broad plateau of the $(n,p)$ excitation function, the NLD effects remaining constant for higher incident energies while the PE contribution increases. 
Nevertheless, even the PE cross sections depend on the $a$ values through the related PLDs \cite{ma98,ah98}, so that the good agreement of our calculated results and the available data does support the present approach.

\begin{figure*} 
\resizebox{2.0\columnwidth}{!}{\includegraphics{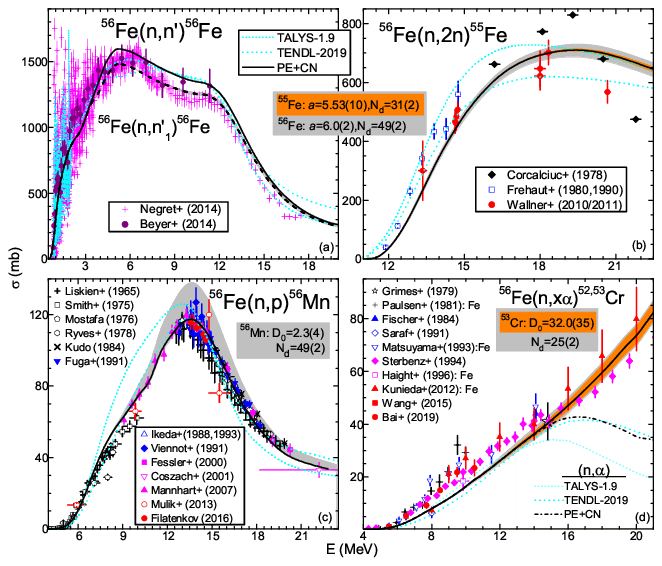}}
\caption{\label{Fig:Fe56nx} As Fig.~\ref{Fig:Fe54nx} but for $^{56}$Fe target nucleus, additionally with (a) inelastic scattering, total (solid curve) and for the first excited at 847 keV (dashed curve), (d) use of only $\alpha$-particle OMP \cite{va14} within both HF+GDH model calculations of the $(n,\alpha)$ reaction (short-dash-dotted curve) and total $\alpha$ emission (solid curve).}
\end{figure*}

{\it The $(n,2n)$ reaction} excitation function shown in Fig.~\ref{Fig:Fe54nx}(b) has firstly been affected by the NLD of the residual nucleus $^{54}$Fe, along its increasing side. 
The level density parameter $a$ of this nucleus, with no resonance data, has also been fixed by the smooth-curve method \cite{chj77}, with an accuracy that leads to uncertainties of calculated cross sections close to 10\%. 

Once the continuum of the final residual nucleus $^{53}$Fe starts to be populated at higher incident energies, it becomes visible an effects mixture of the two NLDs. 
While the NLD uncertainty of only $^{53}$Fe has led to one of $\sim$7\% for the cross sections at the top of the $(n,2n)$ excitation function, the addition of that for $^{54}$Fe increased the one for cross sections up to $\sim$18\%. 
Unfortunately, the discrepancy between the only two data sets as well as the TALYS and TENDL results is even higher at incident energies above 19 MeV. 
Nevertheless, the rest of more recent data are well described by the results of the present work as well as all data by taking into account the NLD total uncertainties. 

{\it The $(n,\alpha)$ reaction} analysis has taken the advantage of confidence in the above-proved suitable account of the main nucleon-emission channels. 
Unfortunately, the HF+PE results of this work are underestimating notably the recent accurate data \cite{zw15,tk18,hb19} as shown in Fig.~\ref{Fig:Fe54nx}(c) below an incident energy of 11 MeV. 
On the contrary, there is also an overestimation above $\sim$17 MeV but with respect to only one earlier data set.

The residual nucleus $^{51}$Cr has the advantage of known resonance data \cite{ripl3}. 
Taking into account the error limits of a medium value of its $s$-wave nucleon-resonance spacings $D_0^{\it exp}$ (Table~\ref{densp}) as well as an uncertainty of 2 levels for the low-lying level number $N_d$, it results an (orange) uncertainty band of the calculated $(n,\alpha)$ cross sections that becomes significant at incident energies higher than 12--13 MeV. 
A larger one corresponds to the assummed limits of the GDH $\alpha$-particle pre-formation probability $\varphi$=0.20$\pm$0.05 [gray band in Fig.~\ref{Fig:Fe54nx}(c)] that increases with energy and covers indeed all data but only above 14--15 MeV. 
Therefore, the most recent and precise $(n,\alpha)$ data remain truly underpredicted just within the energy range where the $\alpha$-particle OMP is the main HF parameter.

On the other hand, the $\alpha$-particle OMP \cite{va14} that was used within both HF and PE model calculations, has firstly been replaced by \cite{va94} for calculation of $\alpha$-particle transmission coefficients involved in HF calculations. 
Then the same replacement concerned also the corresponding PE intranuclear transition rates within the generalized GDH model \cite{ma90}. 
The former replacement provides the agreement already found below $\sim$10 MeV \cite{va94} but an overestimation over 20\% around the incident energy of 16 MeV. 
Actually this enlargement is twice that corresponding to the upper limit of $\varphi$, becoming similarly with the PE increase , i.e. above 18 MeV.
The latter replacement had no effect at the lower energies, where PE mechanism is not yet effective and even not considered in Ref. \cite{va94} that concerned incident energies $\leq$10 MeV.
However, a significant cross-section increase follows at higher energies, becoming even twice the former for the excitation function maximum at $\sim$16 MeV

Therefore, a consistent analysis over the whole energy range of this $(n,\alpha)$ excitation function proves substantial underestimation by $\alpha$-particle OMP \cite{va14} that is removed by the OMP \cite{va94} at the price of a large overestimation for the maximum at $\sim$16 MeV. 
It should be noted that the latter disagreement can not be compensated by either NLD or PE effects within consistent limits.

\subsubsection{$^{57}$Fe de--excitation} \label{56Fe}

{\it The inelastic scattering} of neutrons on $^{56}$Fe was the object of extensive studies while there are especially more recent and high-resolution measurements \cite{rb14,an14,rb18a} which should be considered within any further model analysis. 
This issue is particularly essential for the present work due to the modified neutron OMP (Sec.~\ref{OMPn}) whereas the recent studies used the Koning--Delaroche \cite{KD03} potential. 

The recent total inelastic neutron-scattering data are compared in Fig.~\ref{Fig:Fe56nx}(a) with the latest evaluation \cite{TENDL} and calcutions with TALYS-1.9 default options and present work. 
There are shown also our results for the population of the first excited at 847 keV since its measured $\gamma$-decay was used together with TALYS predictions to derive the total neutron-scattering cross sections \cite{an14}. 
The calculated values of this ratio are decreasing from 1, at the incident energies of 3 MeV, to $\sim$0.92 between the excitation function top around 5 MeV and $\sim$11 MeV, and around 0.95 above 13 MeV. 

Actually, the agreement of our calculated results and the recent data shown in Fig. 12 of Beyer {\it et al.} \cite{rb18a} is better along the increasing side of this excitation function, due to a correction for the $\gamma$-ray angular distribution that causes a reduction of the available data \cite{rb14} by up to 30\%. 
On the other hand, the larger calculated cross sections at higher energies are not related to NLD effects of $^{56}$Fe nucleus, which correspond to an uncertainty band within $\leq$4\%.

{\it The $(n,2n)$ reaction} excitation function shown in Fig.~\ref{Fig:Fe56nx}(b) proves a good agreement with the newest measured data except the last point just above 20 MeV. 
The fast decrease with energy of the two data sets available in this energy range data seems however unphysical.
The NLD effects for both the residual nucleus $^{55}$Fe and $^{56}$Fe, in addition and also shown for the inelastic scattering on the first excited state in Fig.~\ref{Fig:Fe56nx}(a), are below 4\%. 

{\it The $(n,p)$ reaction} excitation function is shown in Fig.~\ref{Fig:Fe56nx}(c), with so much changed behavior regarding the same reaction on $^{54}$Fe.
This is due to the magic $N$ number of the lighter isotope as well as the isotopic effect of $(n,p)$ and $(n,\alpha)$ reaction cross-sections, of decreasing with the isotope mass increase \cite{nim77}. 
On the other hand, it has most recently been proved by Nobre {\it et al.} \cite{gpan20} as an ideal mechanism to probe the NLD of $^{56}$Fe and $^{56}$Mn. 
Their starting point has been that NLD parameters which describe reasonably well $N_d$ and $D_0^{\it exp}$ data do not necessarily lead to consistent calculated cross-section agreements at the precision level required in evaluations.
They finally established additional experimental constrains on NLD through quantitative correlations between reaction cross sections and NLD. 

The uncertainty bands in Fig.~\ref{Fig:Fe56nx} (c) go along the same line as \cite{gpan20}. 
We have also paid due attention to the suitable fit of $N_d$ and $D_0^{\it exp}$ data, however with higher fitted $N_d$ (Table~\ref{densp}) but also close to numbers of levels considered within complete level schemes in RIPL-3 \cite{ripl3}. 
Then, while the cross-section uncertainty bands corresponding to the NLD uncertainties of $^{55,56}$Fe play a minor role in the understanding of eventual lower agreement  between measured and modeling $(n,n')$ data, the $(n,p)$ case is quite different.  
The NLD uncertainty for $^{56}$Mn had indeed no effect at incident energies $\leq$9 MeV, whereas the agreement of data and model results has confirmed the related low-lying level schemes and nucleon OMPs. 

Next, this uncertainty band becomes around or even larger than 15\%, i.e. more than 5 times the experimental errors around the excitation-function maximum. 
Moreover, a suitable agreement with recent data has been obtained using NLD parameters  that correspond to a $D_0$ value close to the higher limit of $D_0^{\it exp}$.
Thus, only a rather low NLD, yet compatible with this experimental data, has been confirmed by analysis of the measured reaction cross sections.

{\it The $\alpha$-particle emission} excitation function analysis, in the absence of available data only for $(n,\alpha)$ reaction, concerns more data sets corresponding to $^{nat}$Fe target. 
However, the data for incident energies $<$15 MeV stand for $(n,\alpha)$ reaction, and include the recent work that have risen questions yet unanswered \cite{zw15,hb19}. 

While this analysis has followed the above-mentioned check of nucleon OMPs and important reaction channels within its energy range, a large underestimation of the experimental data at neutron energies of 8--12 MeV is apparent in Fig.~\ref{Fig:Fe56nx}(d). 
There is also a lower underestimation around 14.1 MeV \cite{rf84}, in the limit of twice the data standard deviation ($\sigma$). 
The related $\alpha$-emission angular distributions and angle-integrated spectrum are discussed in Sec.~\ref{DI} and the main paper, respectively. 

The uncertainty bands are shown firstly with regard to the $D_0^{\it exp}$ limits, and then in addition to those of $N_d$, fitted to obtain the NLD parameters of the residual nucleus $^{53}$Cr (Table~\ref{densp}).
It is thus obvious that effects of $N_d$ uncertainty exist just above the neutron energy of 10 MeV, and yet minor for other 2--3 MeV. 
Only then would they be similar to data errors but around calculated cross sections which are still well below the measured values. 
The effects due to the $D_0^{\it exp}$ uncertainty become also visible, and either equal with the former above 15 MeV, or dominant around 20 MeV.
Nevertheless, the NLD uncertainty band is matching the measured data at incident energies $>$14 MeV while the case is entirely different at lower energies and needs further attention.

\begin{figure*} 
\resizebox{2.0\columnwidth}{!}{\includegraphics{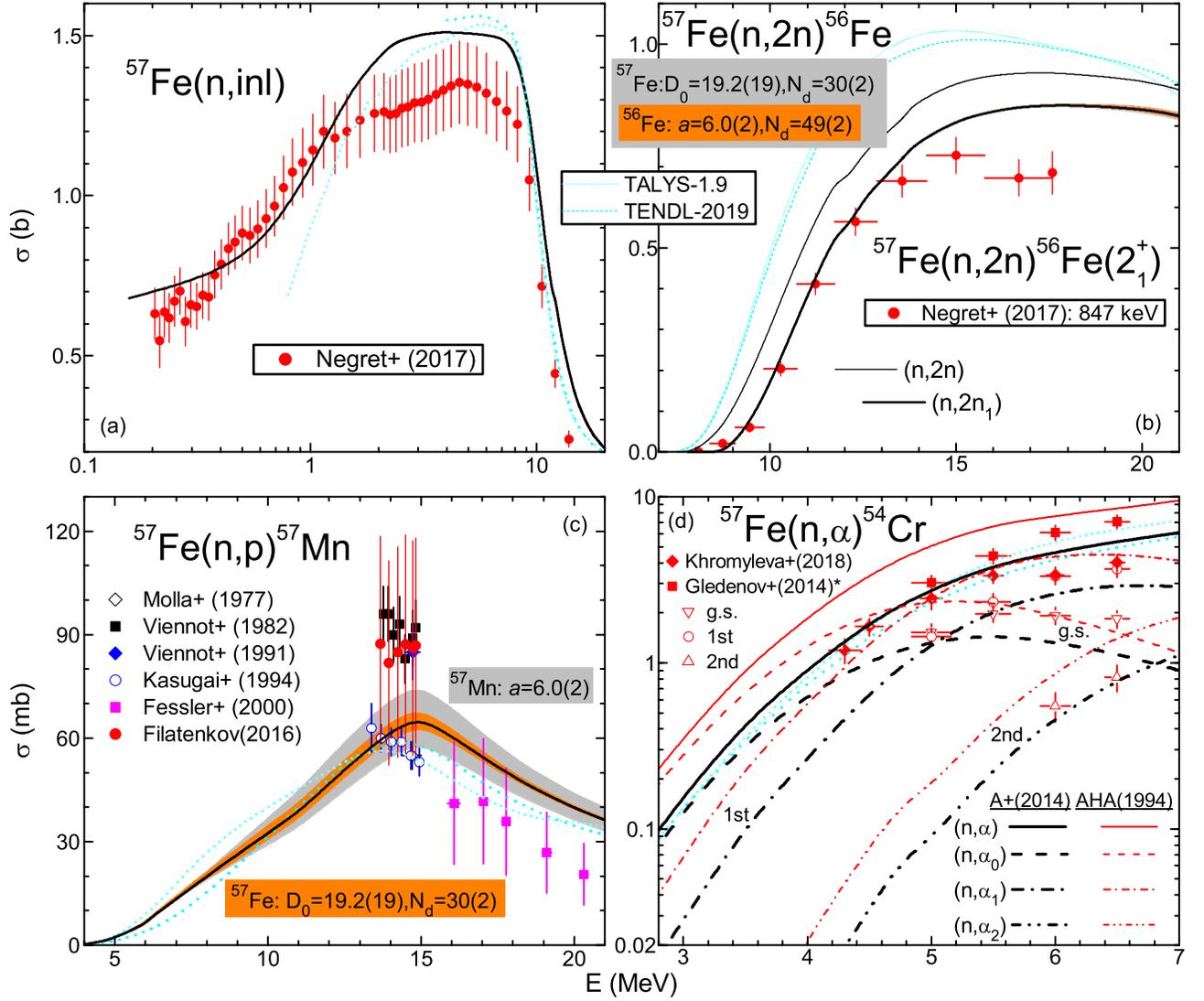}}
\caption{\label{Fig:Fe57nx} As Fig.~\ref{Fig:Fe54nx} but for $^{57}$Fe target nucleus, additionally with (b) total (thin solid curve) and first $2^+$ excited-state population (thick solid state) cross sections of $(n,2n)$ reaction, and (d) total (solid curves) and partial cross sections for population of the ground state (dashed curves), first (dash-dotted curves) and second (dash-dot-dotted curves) excited states, calculated using the $\alpha$-particle OMPs of Rev. \cite{va14} (thick curves) and  Rev. \cite{va94} (thin curves), respectively; total $(n,\alpha)$ cross sections for Gledenov$^*$ {\it et al.} \cite{yg14} correspond to the sum of their measured partial data times our calculated ratio of the total cross section to the sum of the three states.}
\end{figure*}

\subsubsection{$^{58}$Fe de--excitation} \label{57Fe}

{\it The inelastic scattering} of neutrons on $^{57}$Fe being recently analyzed to a large extent \cite{an17}, its first discussion is also essential for this work.
We have used in this respect also the modified neutron OMP (Sec.~\ref{OMPn}) for $^{56}$Fe, with the results shown in Fig.~\ref{Fig:Fe57nx}(a). 
The good agreement of our calculated results and measured data around the neutron energy of 1 MeV is most important for the rest of the model analysis, as a validation of the neutron competition within CN de-excitation. 
At higher energies this agreement is in the $2\sigma$ limit. 

{\it The $(n,2n)$ reaction} excitation function provides, on the basis of the same recent and accurate work \cite{an17}, also a good agreement except the latest two points just above 16 MeV [Fig.~\ref{Fig:Fe57nx}(b)]. 
The overestimation of these data are obviously smaller than results of TALYS default results and particularly the latest TENDL-2019 evaluation. 
The NLD effects for the residual nuclei $^{56}$Fe and $^{57}$Fe in addition, also shown for the $(n,2n)$ reaction to the first excited-state population, are even lower than for the same reaction on the even-even nucleus $^{56}$Fe [Fig.~\ref{Fig:Fe56nx}(b)]. 

{\it The $(n,p)$ reaction} excitation function, shown in Fig.~\ref{Fig:Fe57nx}(c), may  additionally confirm it as an ideal mechanism to probe the residual-nuclei NLD \cite{gpan20}. 
The uncertainty bands in Fig.~\ref{Fig:Fe57nx} (c) correspond just to fitted limits of $N_d$ and either $D_0^{\it exp}$ data of $^{57}$Fe, or average $a$-value for $^{57}$Mn. 
They prove the usefulness of resonance data availability, making possible an uncertainty band three times narrower. 
On the other hand one may note that the NLD uncertainties of the two residual nuclei act in opposite ways on the $(n,p)$ reaction cross sections. 
Therefore, possible systematic errors of the NLD parameters may lead to less wide uncertainty bands of the calculated $(n,p)$ cross sections, which remain thus half-way between the more recent data sets with so large error bars. 

{\it The $(n,\alpha)$ reaction} analysis may concern only but essential measured cross sections for populations of the ground state (g.s.) and first two excited states at 0.835 and 1.824 MeV, respectively, of the even-even residual nucleus $^{54}$Cr at neutron energies between 5 and 6.5 MeV \cite{yg14}.
There are also two additional points added quite recently to the corresponding total $(n,\alpha)$ reaction cross sections between 4 and 5 MeV \cite{tk18}, in addition to the former energies. 
Unfortunately, above 5.5 MeV the two data sets are not consistent while the partial data make possible at least their excitation-dependence analysis as shown in Fig.~\ref{Fig:Fe57nx}(d).

The great advantage of these data consists in their model analysis free of any NLD as well as PE effects. 
Once the transmission coefficients for the eventual open reaction channels are already checked, as within present work, the calculated cross sections are actually given by the $\alpha$-particle OMP. 
This is proved also by the calculated fraction of the three partial cross-section sum to the total cross section including the population of the upper states. 
It goes from 98\%, at the incident energy of 5 MeV, to 90\% at 6.5 MeV. 
Actually, we used this ratio to get the total $(n,\alpha)$ cross sections corresponding to the sum of the partial data measured by Gledenov {\it et al.} \cite{yg14} and shown in Fig.~\ref{Fig:Fe57nx}(d).
These values are however lower than those in Table II of \cite{yg14} by only 3--7\%.

Thus we have found that the total as well as partial cross sections are well described up to $\sim$5.5 MeV using the OMP of Rev. \cite{va14}. 
At the neutron energies of 6 and 6.5 MeV there is an agreement only for the partial cross sections for the second excited state, while the other partial data are underestimated. 
The calculated total $(n,\alpha)$ cross sections at these energies are just between the two data sets \cite{yg14,tk18}.
On the other hand, the calculation results using the OMP \cite{va94} are obviously larger by a factor of two, with reference to the measured data and results of using the potential \cite{va14}. 
Then, the measured data \cite{yg14} are just between the calculated results using the  two OMPs \cite{va14,va94}. 
It is thus apparent the additional attention that should concern both measurement and analysis works.

\subsection{$^{60}$Co de-excitation} \label{Co}

\begin{figure} 
\resizebox{1.0\columnwidth}{!}{\includegraphics{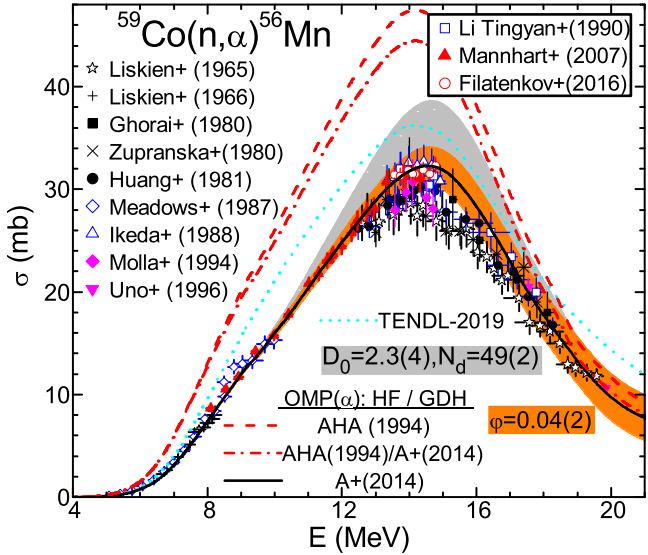}}
\caption{\label{Fig:Co59na0} As Fig.~\ref{Fig:Fe54nx}(c) but for $^{59}$Co target nucleus and reversed colors of the uncertainty bands.}
\end{figure}

Following the above-mentioned detailed study of neutron-induced reactions on $^{59}$Co \cite{vs04}, only an additional analysis of the $\alpha$-particle emission has been carried on in the present work.
It has been motivated by the earlier need to adjust the $\alpha$-particle OMP \cite{va94} by fit of $(n,\alpha)$ measured cross sections within several MeV above the effective reaction threshold.
No such action has presently been necessary, the use of OMP \cite{va14} leading to a good description of entire $(n,\alpha)$ excitation function shown in Fig.~\ref{Fig:Co59na0}. 
This includes the data measured in the meantime at energies where there are no NLD effects illustrated by the corresponding uncertainty band.

On the other hand, the use of the NLD parameters given in Table~\ref{densp} has been of a particular interest because of the same residual nucleus for reactions $^{59}$Co$(n,\alpha)^{56}$Mn and formerly discussed $^{56}$Fe$(n,p)^{56}$Mn. 
The cross-section uncertainty bands corresponding to the NLD effects have similar shapes and the agreement with the measured data is provided by the NLD parameters that correspond to a $D_0$ value close to the higher limit of its experimental value, i.e. NLD values close to the lowest values yet compatible with the resonance data. 

\begin{figure*} 
\resizebox{2.0\columnwidth}{!}{\includegraphics{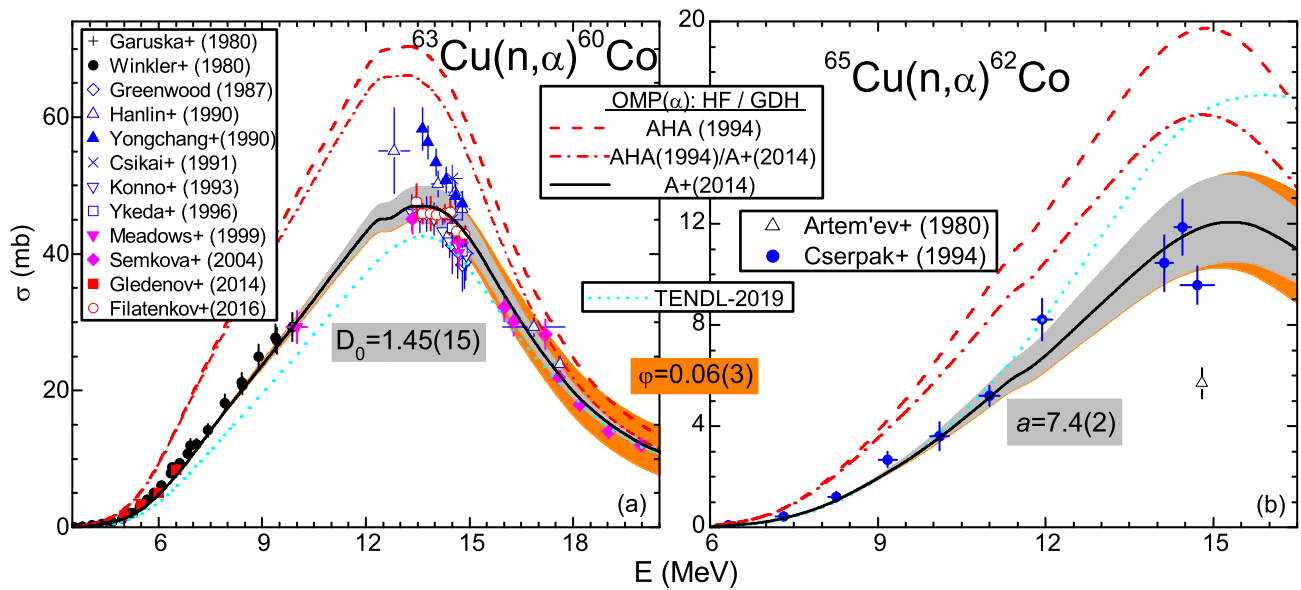}}
\caption{\label{Fig:Cu635na0} As Fig.~\ref{Fig:Co59na0} but for $^{63,65}$Cu target nuclei.}
\end{figure*}

The calculated cross-section uncertainty bands related to the assumed uncertainties of the NLD and PE, respectively, can be also compared in Fig.~\ref{Fig:Co59na0}. 
The latter has been considered to be given by half of the low value $\varphi$=0.04 of the $\alpha$-particle pre-formation probability that correspond to a good agreement of calculated and measured cross sections, $\alpha$-emission angular distributions (Sec.~\ref{DI}), and angle-integrated spectrum \cite{rf86,smg96,k99} (see the main paper). 
It becomes half of the former only around the excitation function maximum, and even larger above the neutron energy of 18 MeV. 
A good agreement there is also at these energies between the calculated and more recently  measured cross sections while the others are within the uncertainty band due to PE effects.

Unlike $^{54}$Fe case, replacement of the $\alpha$-particle OMP \cite{va14} by the earlier one \cite{va94}, firstly in HF calculations and then also within PE approach, has led to a notable overestimation even from the neutron energy of $\sim$6 MeV. 
Only the additional effect within PE account has been smaller due to the much lower $\alpha$-particle pre-formation probability for the $\alpha$-emission from the excited odd-odd nucleus $^{60}$Co. 

Actually, the only data set available for the former analysis was overestimated even at that time. 
On the other hand, that analysis concerned firstly the major role of the low-lying discrete levels (Fig. 2 of \cite{va94}), which was not really the case of the residual odd-odd nucleus $^{56}$Mn.
Nevertheless, the good results provided by use of the OMP \cite{va14} even in these conditions should be kept in mind.

\subsection{Cu isotopes de-excitation} \label{Cu}

The $\alpha$-particle emission by de-excitation of the same nucleus excited through neutron-- as well as proton--induced reactions would answer entirely the question \cite{va17} concerning an eventual difference between the OMPs being able to describe them.
Because of no experimental data for such case, the next option has been the same analysis for various isotopes populated by the two distinct reactions. 
Incident either neutrons on Cu or protons on Ni isotopes provides the first object of this comparative study.

\subsubsection{$(n,\alpha)$ reactions: $^{64,66}$Cu de--excitation} \label{Cuna}

The above-mentioned detailed study of neutron-induced reactions on Cu isotopes \cite{ma08} makes useful in the following only an additional analysis of the $\alpha$-emission.
In a similar way to the involvement \cite{vs04} of the earlier OMP \cite{va94} within $(n,\alpha)$ reaction on $^{59}$Co, a decrease of the real well diffuseness \cite{va94} was used in order to obtain the agreement with data even at lower incident energies.

On the contrary, no change of OMP \cite{va14} parameters has been necessary to obtain a good agreement with the data available for both target nuclei $^{63,65}$Cu (Fig.~\ref{Fig:Cu635na0}).
This agreement concerns firstly the feeding of the low-lying levels, within the several MeV above the effective reaction thresholds.
Then, a similar agreement has also been found around the cross-section maxima and  with the data available for $^{63}$Cu at higher energies.

Thus, an uncertainty band corresponding to the error bar of the available $D_0^{\it exp}$ value for the residual nucleus $^{60}$Co (Table~\ref{densp}) has risen up to $\sim$10\% at incident energies of 13--14 MeV. 
It decreases for higher energies but yet in contact with especially recent measured data. 
On the other hand, there is only one measured data set for $^{65}$Cu target nucleus, below the excitation function top, and even somehow scattered. 
The assumed accuracy for the level density parameter $a$, given by the smooth-curve method, has led to a larger uncertainty band of the calculated cross sections, that seems to cover all these data.

The above-mentioned agreement has corresponded to a low value $\varphi$=0.06 of the $\alpha$-particle pre-formation probability, so close to that found also for $^{59}$Co target nucleus. 
We have taken into account again its 50\% variation in order to estimate the calculated data uncertainty band related to PE effects.
It resulted to be rather similar to those related to the NLD effects for both target nuclei  up to $\sim$15 MeV, while the PE effects become dominant at higher energies. 
Nevertheless, the $\alpha$-particle OMP \cite{va14} provides an accurate account of data available for both Cu isotopes, additional data for $^{65}$Cu being eventually quite useful.

\begin{figure} 
\resizebox{1.0\columnwidth}{!}{\includegraphics{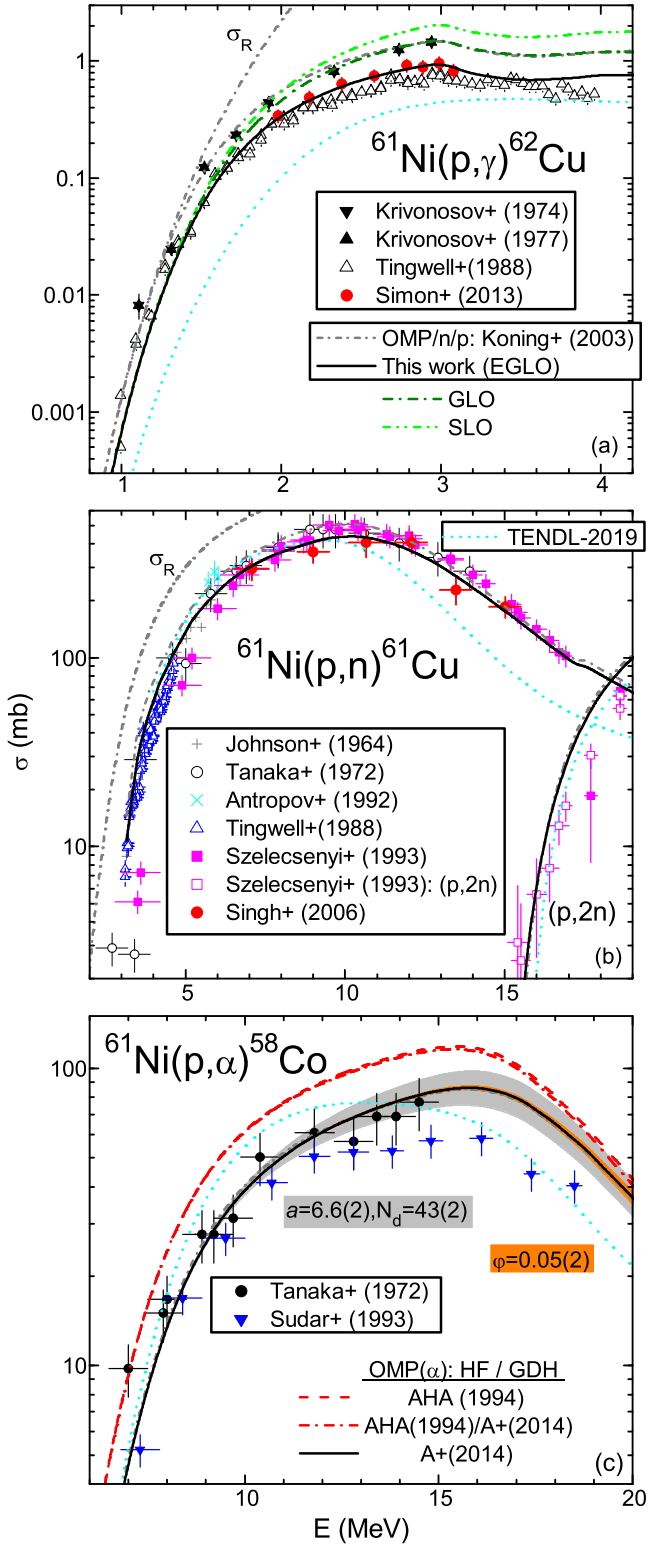}}
\caption{\label{Fig:Ni61px} As Fig.~\ref{Fig:Co59na0} but for proton-induced reactions on $^{61}$Ni, with additional calculated cross sections corresponding to the alternate use of (a) the SLO (dash-dot-dotted curves) and GLO (dash-dotted curves) RSF models for E1 radiations, and (a,b) Koning-Delaroche nucleon OMPs \cite{KD03} (short dash-dotted curves).}
\end{figure}

\subsubsection{$(p,\alpha)$ reactions: $^{62,65}$Cu de--excitation} \label{Nipa}

The key analysis of this section has been that of the reaction $^{64}$Ni$(p,\alpha)^{61}$Co, following the pioneering work of Qaim {\it et al.} \cite{smq95}. 
However, the consistent analysis of all available data for various reaction channels and isotopes has been an essential condition, i.e. including the target nucleus $^{61}$Ni.

\bigskip
\begin{center}
{\it $^{62}$Cu de--excitation} \label{61Nipa}
\end{center}
\bigskip

{\it The $(p,\gamma)$ reaction} analysis makes possible a confirmation of nucleon OMPs mentioned in Sec.~\ref{OMP}. 
Thus, one may see in Fig.~\ref{Fig:Ni61px}(a) that the proton-capture cross sections at the lowest proton energies $\leq$1.3 MeV, where they are equal to $\sigma_R$, are smaller by even a factor of 2 if the modified proton OMP \cite{KD03} is taken into account.
Then, there are rather equal proton-OMP and GLO-model RSF effects, at incident energies just below 3 MeV, where the neutron-emission is not yet open.
While these effects have led to a cross-section increase higher than 50\%, the use of the SLO-model RSF may add another similar increase, in close relation to the RSFs behavior in Fig.~\ref{Fig:RSF_Cu6265}.
Fortunately there are rather recent data which support well the present results.

{\it The $(p,n)$ reaction} measured data are most important for validation of the proton OMP. 
A better agreement especially with the more recent data in Fig.~\ref{Fig:Ni61px}(b) is given by a 10--20\% decrease of the calculated cross sections following the use of modified OMP \cite{KD03}.
It is thus increased the confidence in the presently calculated cross sections. 

{\it The $(p,\alpha)$ reaction} is characterized by the two measured data sets shown Fig.~\ref{Fig:Ni61px}(c).
They agree in the limits of rather large error bars while the calculated cross sections using the $\alpha$-particle OMP \cite{va14} are in between them up to the proton energy of $\sim$12 MeV.
There are, yet minor, changes due to the proton OMP, with no effect on the comparison of calculated and measured data.

At higher energies the agreement remains only with the larger measured data even taking into account the uncertainty band due to the assumed accuracy of $N_d$ and $a$ value obtained by the smooth-curve method for the residual nucleus $^{58}$Co. 
This band has a width up to $\sim$25\% of calculated results, being yet around twice that due to PE effects at the proton energy of 20 MeV.
The latest has been assumed for a change close to half of the $\alpha$-particle pre-formation probability $\varphi$=0.05. 
The so low $\varphi$ value, actually just in between those for the neutron-induced reactions on $^{59}$Co and $^{63,65}$Cu, shows that the larger calculated cross sections at proton energies $\geq$15 MeV are not due to PE contribution.
Maybe additional measurements at these energies could provide a better understanding of this reaction.

Nevertheless, because there are neither NLD nor PE effects below the proton energy of $\sim$10 MeV, the validation of the $\alpha$-particle OMP \cite{va14} is obvious. 
An alternate consideration of the OMP \cite{va94} increases the calculated cross sections by $\sim$50\% at above-mentioned proton energy of $\sim$10 MeV and yet around 35\% at the excitation-function maximum. 
On the other hand, one may note that this $\alpha$-particle OMP effect match at higher energies  the uncertainty band due to NLD accuracy.
So, a real mixture of parameter uncertainties may occur above the incident energy of 20 MeV, making possible meaningful conclusions on them only at lower energies.

\bigskip
\begin{center}
{\it $^{65}$Cu de--excitation} \label{64Nipa}
\end{center}
\bigskip

\begin{figure} 
\resizebox{1.0\columnwidth}{!}{\includegraphics{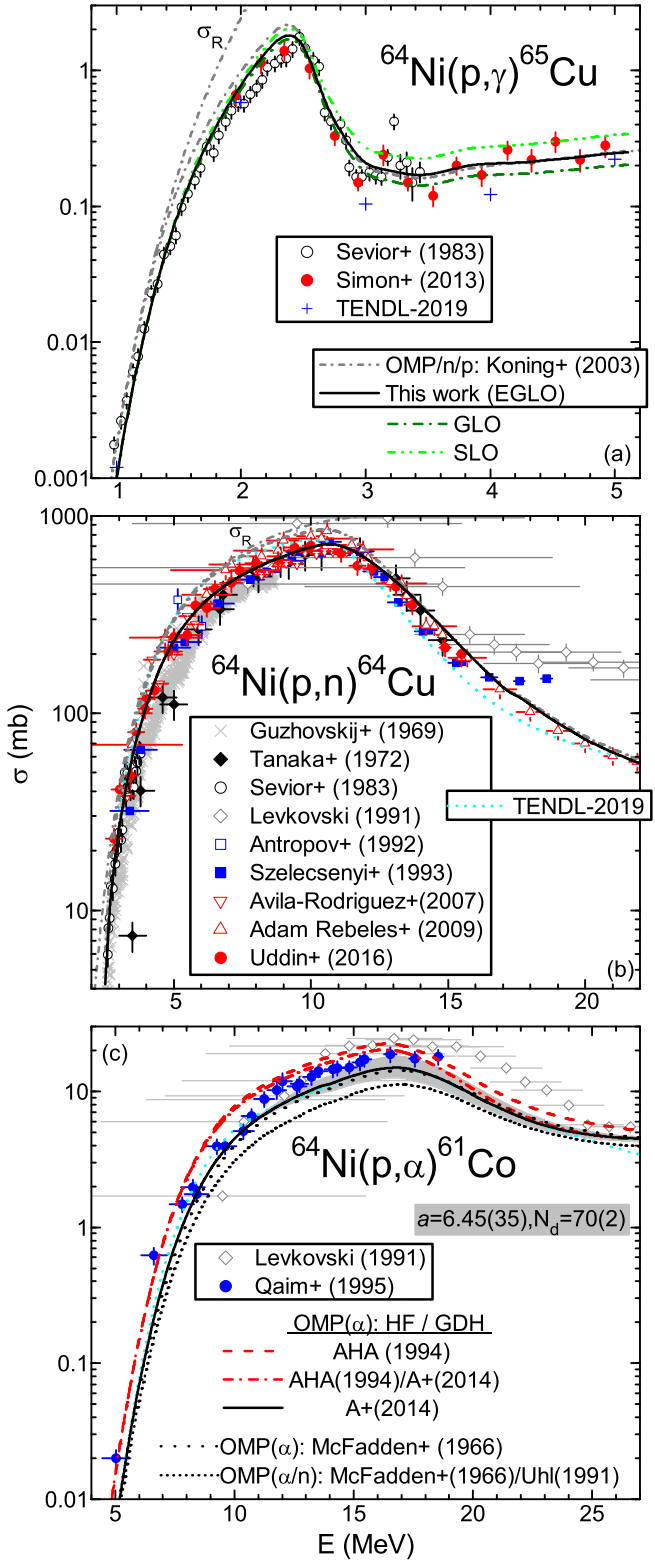}}
\caption{\label{Fig:Ni64px} As Fig.~\ref{Fig:Ni61px} but for $^{64}$Ni target nucleus, with (c) additional calculated cross sections corresponding to the alternate use of the $\alpha$-particle OMP of Ref. \cite{lmf66} (dotted curve), and in addition the neutron OMP \cite{mu91} (short-dotted curve).}
\end{figure}

{\it The $(p,\gamma)$ reaction} analysis presents interesting similarities as well as differences
from the same reaction on $^{61}$Ni.
First, the same is the role of the modified proton OMP \cite{KD03} on proton-capture cross sections at incident energies just below $\sim$2.4 MeV, where the neutron-emission is not yet open. 
On the other hand, the already-discussed quite different RSFs for $^{62,65}$Cu provide a fully changed case for higher energies as shown in Fig.~\ref{Fig:Ni64px}(a).

The RSF values given by the EGLO model in between the lower GLO and higher SLO ones, above the $\gamma$-ray energy of $\sim$2 MeV, provide a similar order of the corresponding capture cross sections. 
A better agreement with the most recent data is given by the GLO model below the proton energy of $\sim$3 MeV, i.e. around the excitation function maximum. 
The differences between the calculated capture cross sections could be considered as a calculation uncertainty band yet within less than 10\%.
 
Moreover, above this maximum, the best average agreement corresponds to the EGLO-model RSF. 
Because of the spreading of also recently measured data, a real uncertainty band due to the RSF models may have however the limits corresponding to the SLO (upper) and GLO (lower) models. 
Its relative width goes over 50\%, rather similar to the more recent data assembly.

{\it The $(p,n)$ reaction} comparative analysis of the measured and calculated data in Fig.~\ref{Fig:Ni64px}(b) is again quite similar to that for $^{61}$Ni target nucleus, taking additionally the advantage of larger reaction cross section.
The more recent data at proton energies above 20 MeV provide also an increased validation of the proton OMP as well as the PE account.

{\it The $(p,\alpha)$ reaction} is also characterized by the two measured data sets,  shown in Fig.~\ref{Fig:Ni64px}(c), of which only one has proper incident-energy error bars. 
It made however the object of the deeper analysis by Qaim {\it et al.} \cite{smq95},  whose review has been challenging in the light of progress in the field, in meantime.   

First, by using the $\alpha$-particle OMP \cite{va14}, we found an obvious underestimation also at lower incident energies but particularly of 20--25\% above 11 MeV.
On the other hand, these calculated cross sections are yet higher by $\sim$50\% than Qaim {\it et al.} HF+PE results. 
In order to understand this difference, we have looked to eventual effects of the OMPs concerned within Ref. \cite{smq95}. 
Thus, use of the $\alpha$-particle OMP of Ref. \cite{lmf66} led to decrease of our results by less than 10\%.
However, the additional use of the neutron OMP \cite{mu91} provides an excitation function that is indeed rather close to the HF+PE results of Qaim {\it et al.} \cite{smq95}. 

The eventually remaining dissimilarity between the HF+PE results of Ref. \cite{smq95} and ours, using their OMPs, can be well motivated by that of the NLDs taken into account. 
Despite the use of the same BSFG model, the NLD parameters for the residual nucleus $^{61}$Co have been based on systematics, in the absence of corresponding resonance data.  
Thus, in Fig.~\ref{Fig:Ni64px}(c) is also shown the uncertainty band of our calculated results, due to the limits assumed for the $N_d$ and $a$ value of $^{61}$Co (Table~\ref{densp}).
It proves that there is no NLD effect up to $\sim$12 MeV, while then its width is close to 40\% of the excitation-function maximum.

Nevertheless, even if this uncertainty band comes near the largest cross sections, there is still an apparent underestimation. 
At the same time, similarly to the case of $^{61}$Ni target nucleus, PE effects become visible around $\sim$12 MeV.
They become only around half of the NLD ones at the cross-sections maximum, and rather equal around 20 MeV.

Moreover, an alternate consideration of the OMP \cite{va94} increases the question marks. 
Thus, it provides calculated cross sections larger than the measured data \cite{smq95} in the $2\sigma$ limit.
Therefore, further consideration of additional pickup direct interaction leading to an increase of the $\alpha$-emission data beyond the HF+PE predictions \cite{smq95} should be carefully taken into account in order to establish the correct $\alpha$-particle OMP. 

\subsection{Zn isotopes de--excitation} \label{Zn}

The analysis of $\alpha$-particle emission in neutron-- as well as proton--induced reactions has the advantage, for de-excitation of Zn isotopes, of accurate data at low incident energies \cite{gz10a,gz10,tk18,jy03,gz07,gz08}.
The $\alpha$-particle OMP study becomes thus less dependent by NLD and PE effects.
Therefore, incident either neutrons on Zn or protons on Cu isotopes provides the second object of this comparative study.

Moreover, the recent analysis of $\alpha$-particle induced reactions on Ni isotopes around the Coulomb barrier \cite{va16} has just proved their suitable description by means of the $\alpha$-particle OMP \cite{va14}. 
Therefore it is so motivating to check the accuracy of the same potential in the inverse reactions. 
At the same time, there are useful also in the present work the corresponding consistent parameter set as well as several competing reactions being already discussed therein Ref. \cite{va16}.

\subsubsection{$(n,\alpha)$ reactions: $^{65,68}$Zn de--excitation} \label{Znna}

The rather recent measured cross sections of the $(n,\alpha)$ reactions on $^{64,67}$Zn, beyond the largest usefulness of quite low incident energies, have the advantage of data available for other reaction channels.
The former analysis of these data makes possible a valuable validation of their common model parameters, to be firstly pointed out in the following.

\bigskip
\begin{center}
{\it $^{65}$Zn de--excitation}
\end{center}
\bigskip

\begin{figure} 
\resizebox{1.0\columnwidth}{!}{\includegraphics{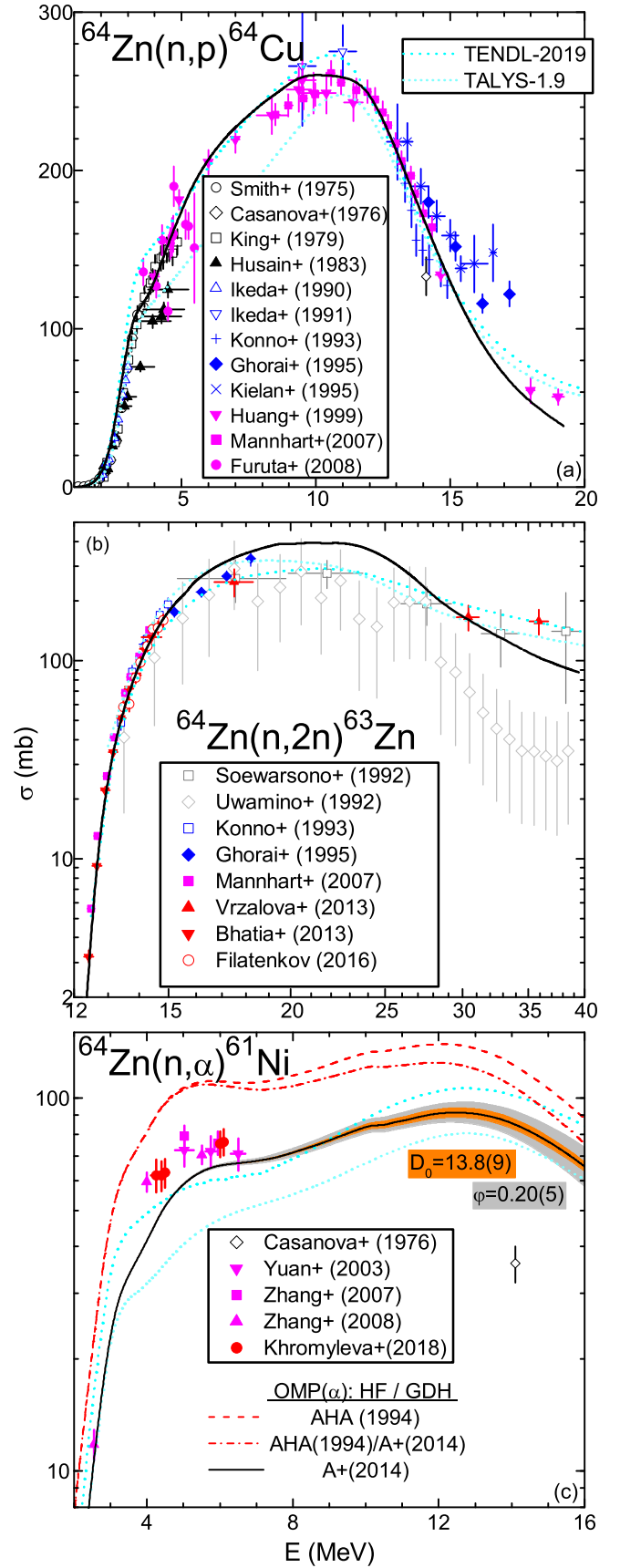}}
\caption{\label{Fig:Zn64nx} As Fig.~\ref{Fig:Fe54nx} but for $^{64}$Zn target nucleus.}
\end{figure}

{\it The $(n,p)$ reaction} comparative analysis of the measured and calculated data in Fig.~\ref{Fig:Zn64nx}(a), making use of more recent measurements, provide again an increased validation of the proton OMP as well as the PE account. 
It seems that the worse description of more recent data above the incident energy of 18 MeV is well compensated by the accurate acoount of the the whole rest of the excitation function.

{\it The $(n,2n)$ reaction} analysis shown in Fig.~\ref{Fig:Zn64nx}(b) validates firstly the neutron OMP and then also the PE account, within incident--energy range below 20 MeV. 
While these energies are of first interest for the present work, more data at higher energies would be quite useful for the modeling support.

{\it The $(n,\alpha)$ reaction} is the first one with recent experimental cross sections [Fig.~\ref{Fig:Zn64nx}(c)] corresponding to feeding of only low-lying states. 
An agreement provided by the use of $\alpha$-particle OMP \cite{va14} has been found at the lowest incident energy of $\sim$2.5 MeV, as well as at the higher one of 6.5 MeV \cite{jy03}. 
However, in between there is an underestimation around or even larger than $2\sigma$.
This is certainly not due to NLD effects, the calculated cross-section uncertainty band corresponding to the measured $D_0$ limits for the residual nucleus $^{61}$Ni becoming visible only above 9 MeV.
A similar case is that of the PE taken into account using an $\alpha$-particle pre-formation probability $\varphi$=0.20$\pm$0.05, the related uncertainty band being twice the former but yet not significant at 6.5 MeV.

On the other hand, an alternate use of the OMP \cite{va94} is followed by additional doubts. 
Thus, the corresponding calculated results are two times larger than the measured point at $\sim$2.5 MeV, and then larger by $\sim$50\% from 4 MeV.

\begin{figure} 
\resizebox{1.0\columnwidth}{!}{\includegraphics{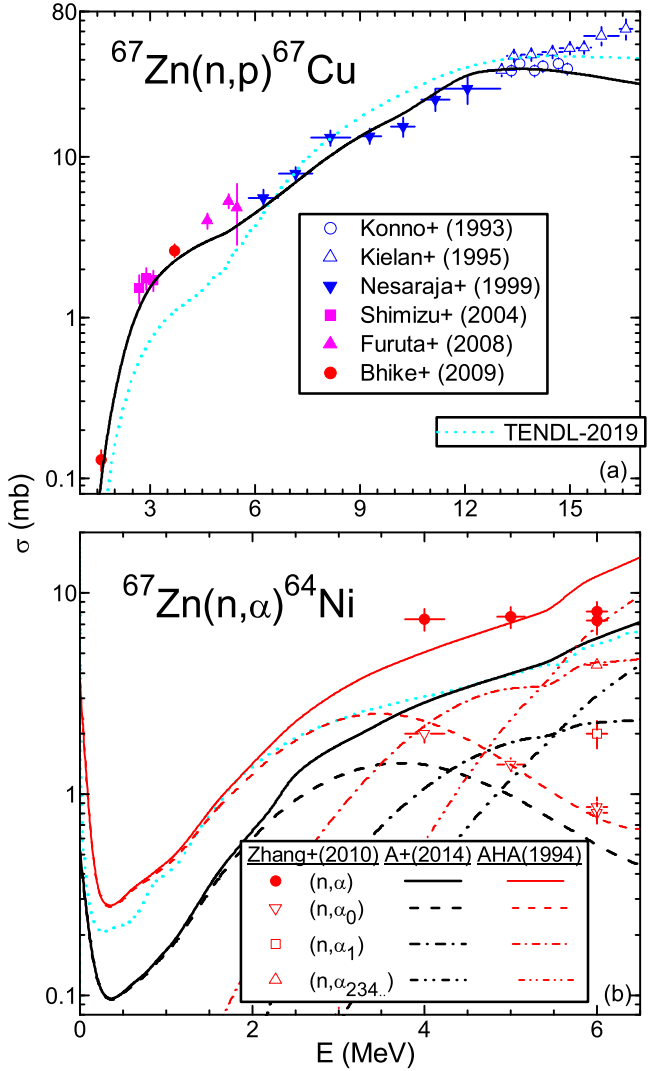}}
\caption{\label{Fig:Zn67na}As Fig.~\ref{Fig:Fe57nx} but for $(n,p)$ and $(n,\alpha)$ reactions on $^{67}$Zn.}
\end{figure}

\bigskip
\begin{center}
{\it $^{68}$Zn de--excitation}
\end{center}
\bigskip

{\it The $(n,p)$ reaction} has fortunately more recent data at incident energies between 1.6 and 6 MeV, as shown in Fig.~\ref{Fig:Zn67na}(a). 
This is the energy range where there are available the $\alpha$-emission data, various model parameters being involved within analysis of both reaction channels.
The agreement of the present model calculations with these data as well as the data available at energies $\leq$14 MeV is quite good, while at higher energies it is so with one of two inconsistent data sets. 

{\it The $(n,\alpha)$ reaction} partial cross sections for the g.s., first excited state at 1.346 MeV, and higher states from 2.277 MeV, were firstly measured at the neutron energy of 6 MeV \cite{gz10a} and, in addition, at 4 and 5 MeV \cite{gz10}.
The usefulness of these data for the assessment of the $\alpha$-particle OMP is that already underlined for the same reaction on $^{57}$Fe \cite{yg14,tk18}. 

The analysis results are similar however only for the g.s. partial cross sections. 
The measured data trend is well described by calculations using both OMPs \cite{va14,va94}, the former potential leading to an underestimation of $\sim$30\%. 
The usual greater values given by the latter potential \cite{va94} are in agreement with the experimental values in this case. 
It seems however that their decrease with the incident--energy increase is faster than either the measured data or results given by the former OMP \cite{va14}.

Nevertheless, more ambiguous is the excitation function slope between 4 and 6 MeV, given by $\alpha$-particle OMPs \cite{va14,va94} at variance with the experimental one. 
The same is true also for the TENDL-2019 evaluation, which actually used the former potential. 
Therefore, additional work should concern this issue, maybe also experimentally.

\subsubsection{$(p,\alpha)$ reactions: $^{64,66}$Zn de--excitation} \label{Znpa}

\begin{figure*} 
\resizebox{1.8\columnwidth}{!}{\includegraphics{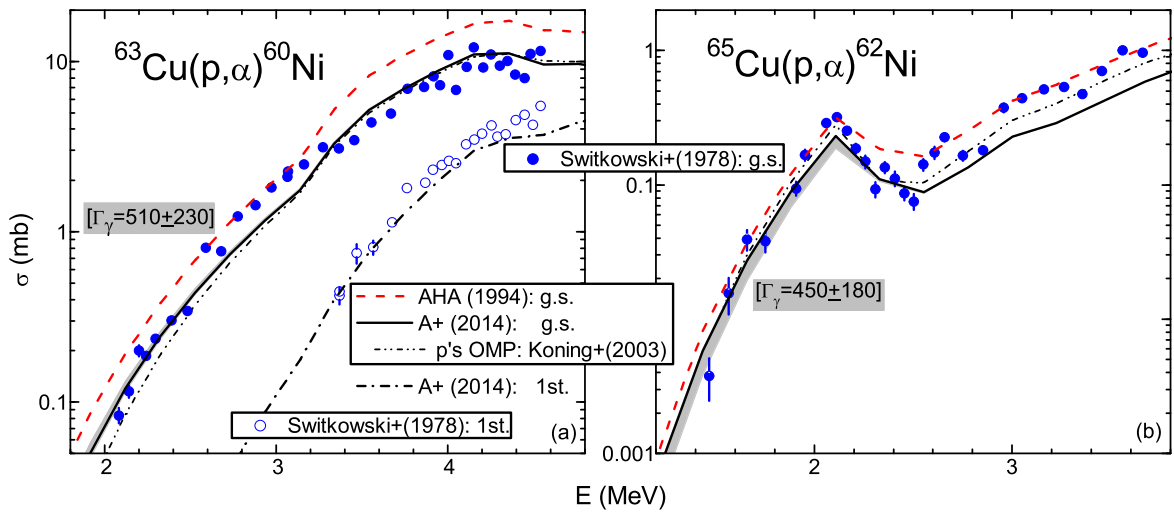}}
\caption{\label{Fig:Cu635pa} (Color online) Comparison of measured \cite{zes78} and calculated partial cross sections of $(p,\alpha)$ reaction on $^{63,65}$Cu and population of g.s. (solid curves) and first excited state (dash-dotted curve), using the $\alpha$-particle OMP of Ref. \cite{va14}, and the OMPs alternate use of either Ref. \cite{va94} for $\alpha$-particles (dashed curves), or \cite{KD03} for protons (dash-dot-dotted curves); uncertainty bands correspond to the error bars of $s$-wave average radiation widths $\Gamma_{\gamma}$ (in meV) deduced from systematics \cite{ripl3}.}
\end{figure*}

The analysis of the $(p,\alpha)$ reaction on $^{63,65}$Cu has taken the advantage of the recent analysis of $\alpha$-particle induced reactions on Ni isotopes around the Coulomb barrier \cite{va16}. 
Thus, the corresponding consistent parameter set as well as the $(p,\gamma)$ and $(p,n)$ competing reactions, of interest also in the present work, have already been discussed. 
That is why in the following we have taken into account their suitable description,  formerly proved, and have straightforwardly proceeded to $(p,\alpha)$ analysis. 

Moreover, there are partial cross sections measured earlier \cite{zes78} for g.s. population of the residual even-even nuclei $^{60,62}$Ni and first excited state of $^{60}$Ni, at proton energies below 4.8 MeV (Fig.~\ref{Fig:Cu635pa}). 
The original study of Switkowski {\it et al.} looked for $(p,n)$ threshold effects that are quite different due to a lower threshold of only 2.167 MeV for $^{65}$Cu, vs. 4.215 MeV for $^{63}$Cu. 
They found indeed a severe fall of the $(p,\gamma)$ and $(p,\alpha_0)$ cross sections for $^{65}$Cu, at variance to the case of higher threshold for $^{63}$Cu. 

The previous analysis of proton-induced reactions on $^{63,65}$Cu \cite{va16} describe well the $(p,n)$ threshold effect for the $(p,\gamma)$ reaction on $^{65}$Cu, as well as the importance of using improved proton OMP (Fig. 1 of \cite{va16}). 
In the present work we found firstly worthy of note the effect of the alternate use of the proton OMP of Koning-Delaroche \cite{KD03} and its modified version for low energies. 
Thus, this effect is rather normal only for $^{65}$Cu, the former potential leading to larger $(p,\alpha)$ cross sections above the $(p,n)$ threshold, while it is absent or even inverted just above the $(p,\alpha)$ threshold for $^{63}$Cu. 

A different effect is also shown in Fig.~\ref{Fig:Cu635pa} by the uncertainty band corresponding to the error bars of $s$-wave average radiation widths $\Gamma_{\gamma}$ deduced from systematics \cite{ripl3}. 
It is minor for $^{63}$Cu but similar to the data error bars for $^{65}$Cu below the $(p,n)$ threshold. 
However, a contribution in this respect has also the isotopic effect leading to $(p,\alpha)$ cross sections lower by even an order of magnitude for the heavier isotope.
  
We paid due attention to various effects as above-mentioned in order to avoid any misinterpretation of the rather questionable comparation of the presently calculated cross sections using the $\alpha$-particle OMP \cite{va14}, and the measured data. 
The best agreement has been obtained for the excitation functions of first excited state of $^{60}$Ni as well as $^{60,62}$Ni g.s. within $\sim$1 MeV above the $(p,\alpha)$ threshold. 
There is also a somewhat good account of the sound $(p,n)$ threshold effect for $^{65}$Cu, as well as of the weak one for $^{63}$Cu just above 4.2 MeV.
Otherwise, the $(p,n)$ threshold effect for $^{65}$Cu is followed by a rather constant underestimation. 

On the other hand, the calculated cross sections using the earlier $\alpha$-particle OMP \cite{va94}, larger by $\sim$30\% than the former results, show distinct cases for the two nuclei. 
Thus, except several data points around the proton energy of 3 MeV, they are overestimating the data for $^{63}$Cu but match the data for $^{65}$Cu above the $(p,n)$ threshold effect. 
Unfortunately, even Switkowski {\it et al.} \cite{zes78} considered the resolution of their data to be insufficient for a more detailed analysis.

Nevertheless, this analysis of $(p,\alpha)$ reaction on $^{63,65}$Cu is pointing out that the measured data for the heavier isotope could be also well described by \cite{va14} if an additional contribution would exist. 
This eventual addition would be negligible for $^{63}$Cu, due to the isotopic effect, if it is rather similar for both nuclei and as large as the difference between the measured and calculated values for $^{65}$Cu. 
These issues suggest that a DI contribution is the one missing within modeling work.

\section{Pickup $DI$ modeling} \label{DI}

The due consideration of DI role within the $\alpha$-emission in both neutron-- and proton--induced reactions is obviously leading to an increase beyond the PE+HF predictions.
Moreover, since the beginning of '90s it is concluded that the pickup instead of knockout has the main DI contribution to the low-lying levels in $(p,\alpha)$ and $(n,\alpha)$ reactions (\cite{eg92,eg89,eg91} and Refs. therein). 
While this conclusion was achieved at incident energies above 20 MeV, there were Qaim {\it et al.} \cite{smq95} extending it even around 10 MeV  
However a better connection of their phenomenological results to available spectroscopic data would be useful for an increased predictive power as needed in the present work.

In the present work, the pickup contributions to $(p,\alpha)$ and $(n,\alpha)$ reactions have been determined within the distorted wave Born approximation (DWBA) formalism using the code FRESCO \cite{FRESCO} as well as the same above-mentioned particle OMPs. 
One--step reaction has been considered through the pickup of $^3$H and $^3$He clusters, respectively. 
Moreover, the "spectator model" \cite{smits76,smits79} was involved, where the two transferred either neutrons or protons in $(p,\alpha)$ and $(n,\alpha)$ reactions, respectively, are coupled to zero angular momentum acting as spectators, while the transferred orbital ($L$) and total ($J$) angular momenta are given by the third, unpaired nucleon of the transferred cluster.
The prior form distorted--wave transition amplitudes, and the finite--range interaction have been considered. 
The p-$^3$H as well as n-$^3$He effective interaction in the $\alpha$ particle are assumed to have a Gaussian shape \cite{eg91,eg88,eg88b}:
\begin{equation}\label{eq:1}
V_r=-V_0 e^{-(r/r_0)^2} ,
\end{equation}
\noindent
where $r_0$=2 fm, and $V_0$ is determined by fit of the binding energies of $^3$H and $^3$He, respectively.

The three--nucleon transferred cluster bound states were generated in a Woods--Saxon real potential \cite{smits79,eg88,eg91,eg88b} with the depth adjusted to fit the separation energies in the target nuclei. 
The number of nodes ($N$) in the radial three--nucleon cluster wave function was determined by the harmonic--oscillator energy conservation rule \cite{smits76,smits79}:
\begin{equation}\label{eq:2}
2N+L=\Sigma^{3}_{i=1}[2(n_{i}-1)+l_{i}],
\end{equation}
\noindent
where $n_i$ and $l_i$ are the single--particle shell--model state quantum numbers.
If the three nucleons are picked from $1f2p$ shell, in the present work on A$\sim$60 target nuclei, then $2N+L$=9. 
Otherwise, if the unpaired nucleon is picked from $2s1d$ shell, then $2N+L$=8, while if it is from $1g$ subshell, it follows that $2N+L$=10. 

Nevertheless, the assessment of DI cross sections is subject to available information on spectroscopic factors related to populated states, outgoing particle angular distributions, or at least differential cross--section maximum values.

\subsection{$^{64}$Ni$(p,\alpha)$$^{61}$Co} \label{pa}

Our starting point on the pickup contribution to $\alpha$-emission in nucleon--induced reactions at low energies was the pioneering work of Qaim {\it et al.} \cite{smq95}. 
In order to describe $^{64}$N$(p,\alpha)$$^{61}$Co reaction, they normalized the formerly calculated semimicroscopic pickup contribution to account for the measured data at 15 MeV. 
However, this normalization depends notably on the preceding PE+HF calculated results. 
Since these are quite different as shown in Sec.~\ref{64Nipa}, we have looked for absolute values by DWBA analysis using spectroscopic factors corresponding to the outgoing $\alpha$-particle angular distributions reported at 15 MeV by Jolivette and Browne \cite{plj78}, and at 30 MeV by Smits {\it et al.} \cite{smits79}. 

\begin{figure} 
\resizebox{1.0\columnwidth}{!}{\includegraphics{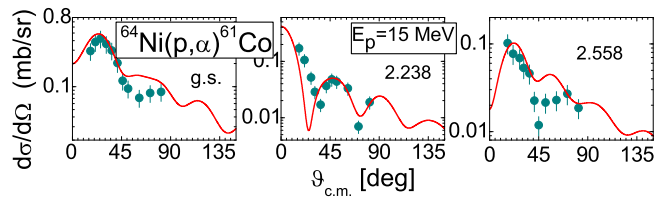}}
\caption{\label{Fig:Ni64pa} (Color online) Comparison of measured (solid circles) \cite{plj78} and calculated (solid curves) $\alpha$-particle angular distributions of $^{64}$Ni$(p,\alpha)^{61}$Co pickup transitions to states, with excitation energies in MeV, at incident energy of 15 MeV.}
\end{figure}

\begin{figure} 
\resizebox{1.0\columnwidth}{!}{\includegraphics{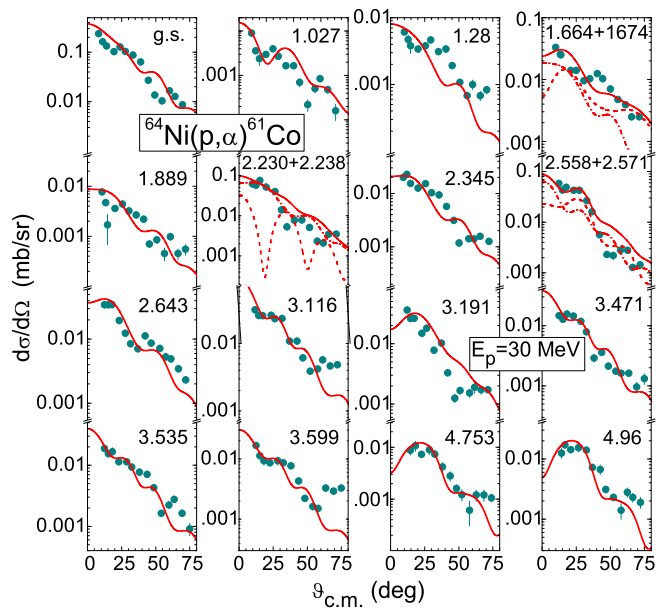}}
\caption{\label{Fig:Ni64pa30} As Fig.~\ref{Fig:Ni64pa} but for incident energy of 30 MeV \cite{smits79}, and additional sum (solid curves) for doublets at 2.238, 2.558, 1.664 MeV (dashed curves), and 2.230, 2.571, 1.674 MeV (dash-dotted curves), respectively.}
\end{figure}

Thus, for description of picked $\alpha$-particle angular distributions (Figs.~\ref{Fig:Ni64pa}-\ref{Fig:Ni64pa30}) within "spectator model" \cite{smits76,smits79}, the 0$^+$ g.s. of $^{64}$Ni target nucleus led to the transferred angular momentum $L$ being fully set by the residual $^{61}$Co final--state spin and parity. 
 
A particular note should concern the angular distributions corresponding to 2.238 and 2.558 excited states of $^{61}$Co residual nucleus.
The spectroscopic factors obtained through their analysis at 15 MeV incident energy (Fig.~\ref{Fig:Ni64pa}) did not describe the data at 30 MeV (Fig.~\ref{Fig:Ni64pa30}). 
It became yet possible to describe the data at 30 MeV by taking into account the excitation of the doublets shown in Fig.~\ref{Fig:Ni64pa30}.
Finally, 19 states \cite{ensdf}, until the excitation energy of $\sim$5 MeV, have been taken into account in the assessment of the pickup excitation function that is discussed in the main paper.

\begin{figure*} 
\resizebox{2.0\columnwidth}{!}{\includegraphics{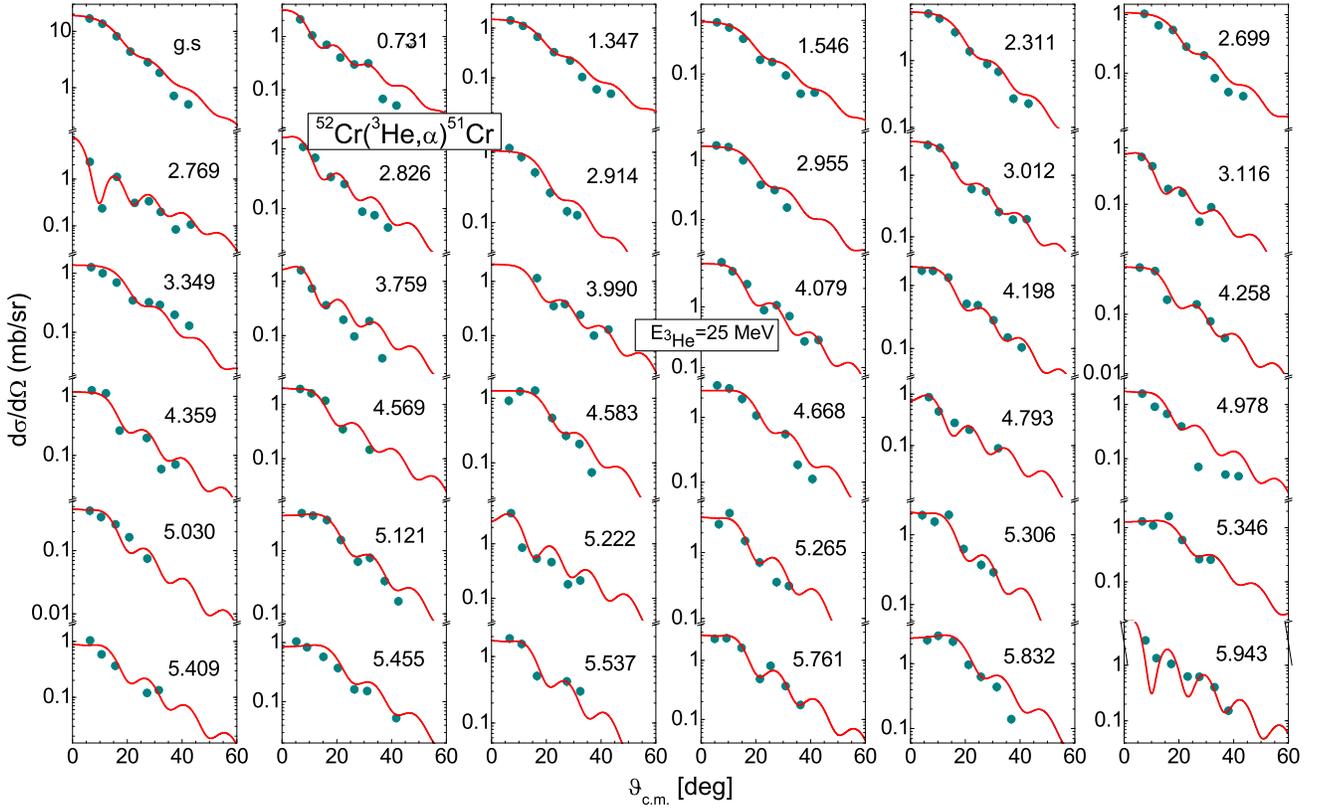}}
\caption{\label{Fig:Cr52He3a}  As Fig.~\ref{Fig:Ni64pa} but for $\alpha$-particle angular distributions of $^{52}$Cr$(^{3}$He,$\alpha)^{51}$Cr pickup transitions, at incident energy of 25 MeV \cite{fortier}.}
\end{figure*}

\subsection{$(n,\alpha)$ reactions} \label{na}
 
There are scarce information with regard to the analysis of pickup $(n,\alpha)$ reactions for A$\sim$60 target nuclei. 
Thus no measured angular distribution of $\alpha$ particles from pickup processes has been found for the $(n,\alpha)$ reactions within the present work. 

Consequently we carried out the pickup $(n,\alpha)$ cross-sections calculations standing on the spectator role of the picked proton pair \cite{eg88,eg88b}, with the spectroscopic factors given by Glendenning (Table II of Ref. \cite{glendenning}).
At the same time, the spectroscopic factors for the picked neutron, that becomes thus responsible for the angular-momentum transfer, have been obtained by angular-distribution analysis for neutron pickup processes, as $(^3He,\alpha)$, $(d,t)$, and $(p,d)$, populating the same  residual nuclei. 

\subsubsection{$^{54}$Fe$(n,\alpha)^{51}$Cr}

The pickup $^{54}$Fe$(n,\alpha)^{51}$Cr cross--section calculation was carried out using the  assumed spectator proton pair. 
Thus, the transferred angular momentum $L$ was uniquely determined by the  residual--nucleus final state.
The Glendenning spectroscopic factor \cite{glendenning} corresponding to the transferred spectator proton pair from the $1f_{7/2}$ subshell was involved in the DWBA analysis too.

The picked-neutron spectroscopic factors were obtained from the comparison of the measured $\alpha$-particle angular distributions \cite{fortier} and the DWBA analysis of $^{54}$Fe($^{3}$He,$\alpha$)$^{51}$Cr reaction, at 25 MeV incident energy, (Fig.~\ref{Fig:Cr52He3a}).  
Thus, 36 excited states up to the excitation energy of 5.943 MeV \cite{fortier,NDS51} have been considered for calculation of $^{54}$Fe$(n,\alpha)^{51}$Cr pickup excitation function in the main paper.  

\subsubsection{$^{56}$Fe$(n,\alpha)^{53}$Cr}

\begin{figure} 
\resizebox{1.0\columnwidth}{!}{\includegraphics{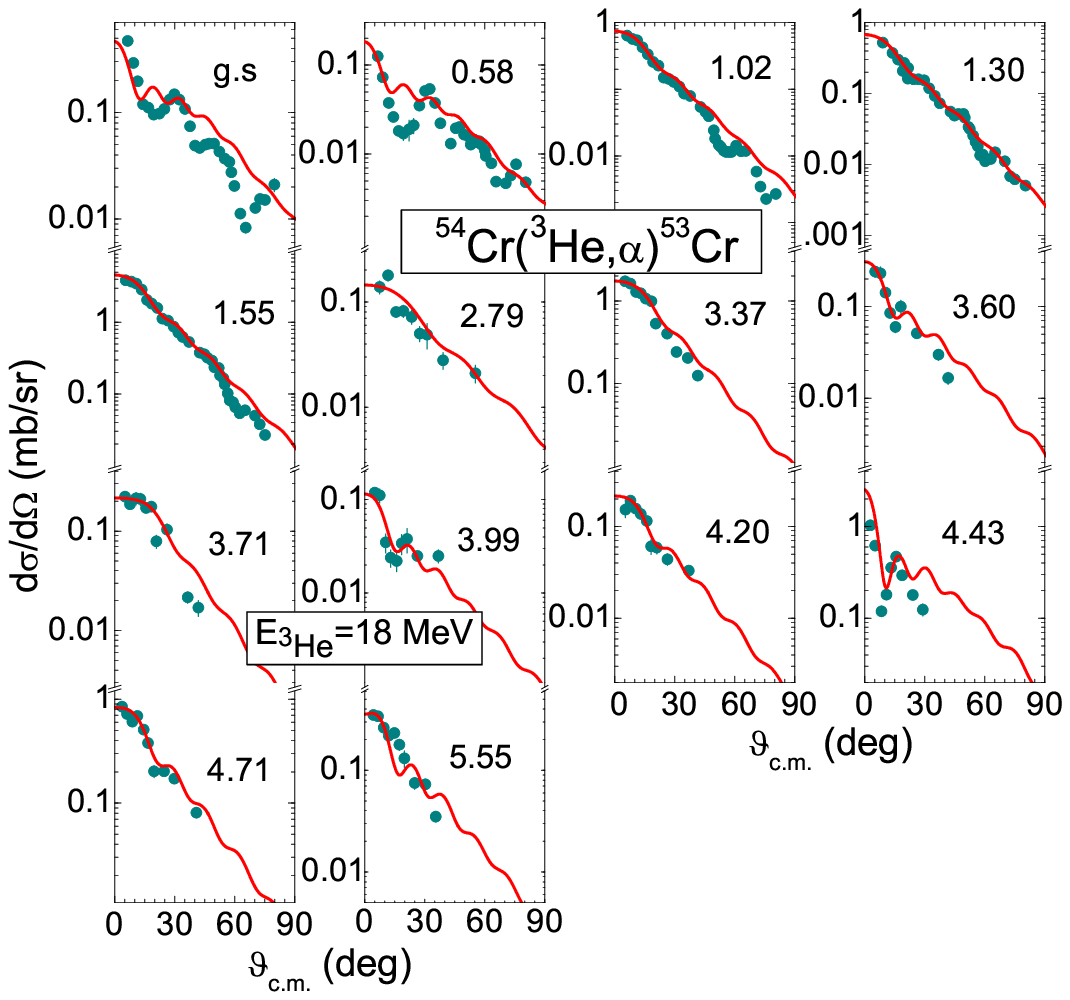}}
\caption{\label{Fig:Cr54He3a} As Fig.~\ref{Fig:Ni64pa} but for $\alpha$-particle angular distributions of $^{54}$Cr$(^3He,\alpha)^{53}$Cr pickup transitions, at incident energy of 18 MeV \cite{david}.}
\end{figure}

\begin{figure} 
\resizebox{0.8\columnwidth}{!}{\includegraphics{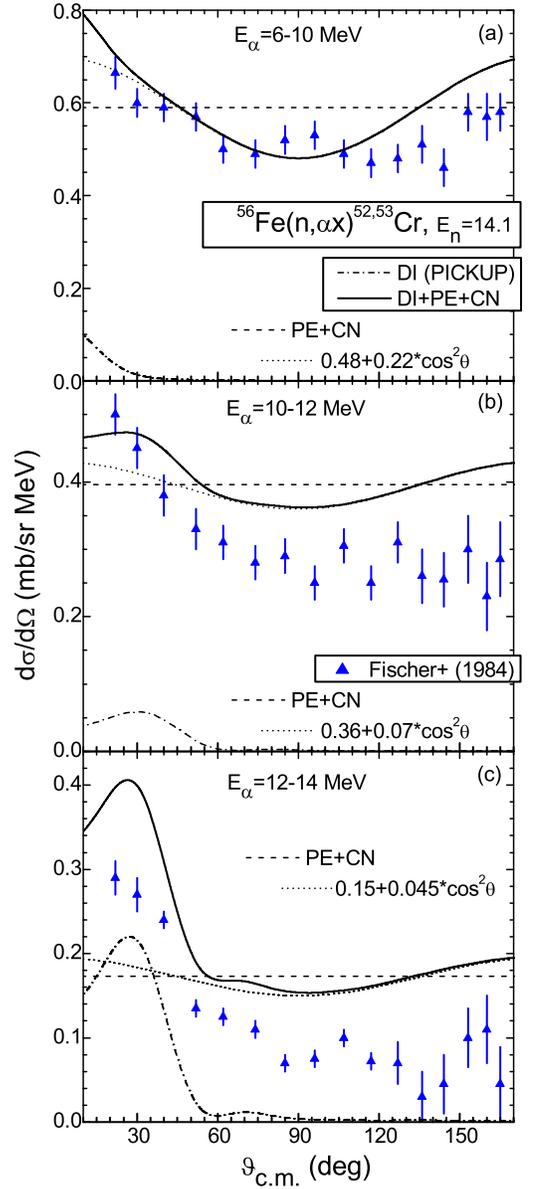}}
\caption{\label{Fig:ad56Fe141} (Color online) Comparison of measured \cite{rf84} angular distributions for $^{56}$Fe$(n,\alpha)$ reaction and different $\alpha$-energy bins in the c.m. system, of (a) 6-10 MeV, (b) 10-12 MeV, and (c) 12-14 MeV, and calculated values of the pickup DI (dash-dotted curves), PE+CN (dashed curves) equivalent forms $a$+$b$$\cdot$$cos^2$$\theta$ (dotted curves), and sum of the DI+PE+CN components (solid curves).}   
\end{figure}

The above-described approach concerned also the pickup $^{56}$Fe$(n,\alpha)^{53}$Cr cross--section calculation, using the assumed spectator proton pair and the transferred angular momentum $L$ determined by the residual state.
Thus, the Glendenning spectroscopic factor \cite{glendenning} corresponding to the transferred spectator proton pair from the $1f_{7/2}$ subshell was also involved in the DWBA analysis too.

The picked-neutron spectroscopic factors were obtained from the comparison of the measured $\alpha$-particle angular distributions \cite{david} and the DWBA results for $^{54}$Cr$(^3He,\alpha)^{53}$Cr reaction, at 30 MeV incident energy, (Fig.~\ref{Fig:Cr54He3a}).  
Then, 18 excited states up to the excitation energy of 5.557 MeV \cite{david,whitten,NDS53} were involved in calculation of $^{56}$Fe$(n,\alpha)^{53}$Cr pickup excitation function (see the main paper). 
The same calculations concerned also the analysis of $\alpha$-emission angular distributions and an angle-integrated spectrum at 14.1 MeV  \cite{rf84} in the main paper. 

Actually, Fischer {\it et al.} \cite{rf84} measured double-differential $\alpha$--emission spectra for 16 reaction angles ranging between 22$^o$--165$^o$, at the incident energy of 14.1 MeV. 
Because of rather large statistical errors, results integrated over either energy or angle were presented, as well as the corresponding total $\alpha$-emission cross section of the $^{56}$Fe$(n,\alpha)$ reaction also shown in Fig.~\ref{Fig:Fe56nx}(d). 
Comparison of calculated angular distributions for the $\alpha$-energy bins, within c.m. system, of 6-10 MeV, 10-12 MeV, and 12-14 MeV, and the Fischer {\it et al.} data is shown in Fig.~\ref{Fig:ad56Fe141}.

The above--mentioned pickup results were added to the equivalent forms $a$+$b$$\cdot$$cos^2$$\theta$ \cite{te58,sc63} corresponding to the PE+CN isotropic component, as shown in Fig.~\ref{Fig:ad56Fe141}. 
First, the isotropic PE+CN component goes from a good agreement with data at lower $\alpha$-particle energies (6--10 MeV), to some overestimation at higher energies (10--12 and 12--14 MeV).
Then, the relevant point is that the anisotropy of the measured distributions is well accounted by the present DI pickup approach. 

\begin{figure} 
\resizebox{0.8\columnwidth}{!}{\includegraphics{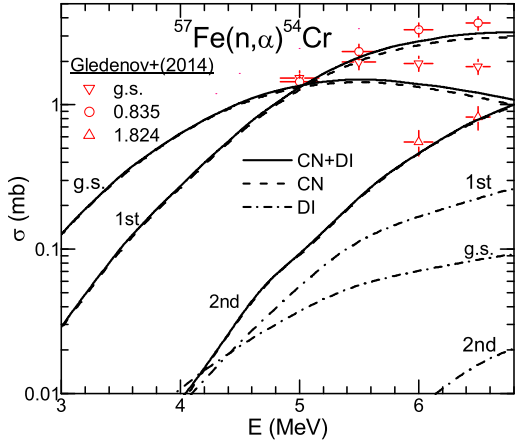}}
\caption{\label{Fig:Fe57na1}  As Fig.~\ref{Fig:Fe57nx}(d) but only for g.s. (firstly upper, then middle curves), first (middle, then upper curves) and second (lowest curves) excited states of $^{54}$Cr, and CN cross sections calculated using the $\alpha$-particle OMPs of Rev. \cite{va14} (dashed curves), DI pickup contributions (dash-dotted curves), and CN+DI sum (solid curves).}
\end{figure}

\begin{figure} 
\resizebox{0.7\columnwidth}{!}{\includegraphics{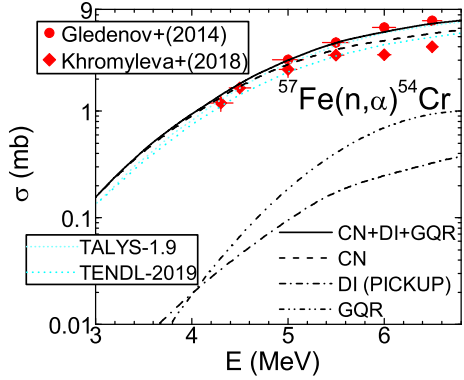}}
\caption{\label{Fig:Fe57na} As Fig.~\ref{Fig:Fe57nx}(d) but only for total $(n,\alpha)$ reaction cross sections \cite{yg14,tk18}, and additional like--GQR component (dash-dot-dotted curve).}
\end{figure}

\subsubsection{$^{57}$Fe$(n,\alpha)^{54}$Cr} 

The lack of $\alpha$-particle angular-distribution data or spectroscopic factors for picked-neutron reactions, e.g., $(p,d)$, $(d,t)$ or $(^3He,\alpha)$ on $^{57}$Fe, makes possible only a qualitative estimation of the related pickup $(n,\alpha)$ cross sections.
We rely on the likeness sequence of the first three low-lying sates of $^{54}$Cr and $^{56}$Fe nuclei, having the same number 30 of neutrons, and 24 and 26, respectively, protons in the $1f_{7/2}$ subshell. 
Thus, we used for this specific reaction the neutron spectroscopic factors reported by Daehnick {\it et al.} \cite{NDS56,daehnickdt} from analysis of $^{57}$Fe$(d,t)^{56}$Fe pickup reaction, and Glendenning spectroscopic factor \cite{glendenning} corresponding to the transferred spectator proton pair from $1f_{7/2}$ subshell. 

The consequently pickup cross sections obtained for the three low-lying sates of $^{54}$Cr by $(n,\alpha)$ reaction on $^{57}$Fe are shown in Fig.~\ref{Fig:Fe57na1} in addition to the CN components given formerly in Fig.~\ref{Fig:Fe57nx}(d). 
While there is indeed an order of magnitude between the two mechanism contributions, a slightly improved trend is provided by the pickup consideration for the g.s. and first excited state. 
Moreover, an agreement seems to become possible in the case of the larger cross sections for $2_1^+$ excited state even at incident energies of 6 and 6.5 MeV. 
It remains thus a significant underestimation at these energies only for the $0^+$ g.s. while a good agreement is already provided for the $4_1^+$ excited state by CN mechanism. 


Following the discussion in the main paper on like--GQR contributions in $(n,\alpha)$ reaction on $^{54,56}$Fe, a similar attempt has concerned also the analysis of these data. 
Thus, their description can be obtained by inclusion of a Gaussian distribution at $E_{GQR}$=16.792 MeV \cite{js81} for the excited nucleus $^{58}$Fe, with a FWHM width of 2.35 MeV and a peak cross section of 1 mb (Fig.~\ref{Fig:Fe57na}). 
It results again a like-GQR component larger than the DI pickup cross sections along the yet increasing side of the former. 
Both of them are still minor but provide an increased agreement with one of the two data sets. 
Nevertheless these data sets, that are not consistent above the incident energy of 6 MeV, make less certain any definite conclusion on a possible like--GQR component within $\alpha$-emission.

\begin{figure} 
\resizebox{0.8\columnwidth}{!}{\includegraphics{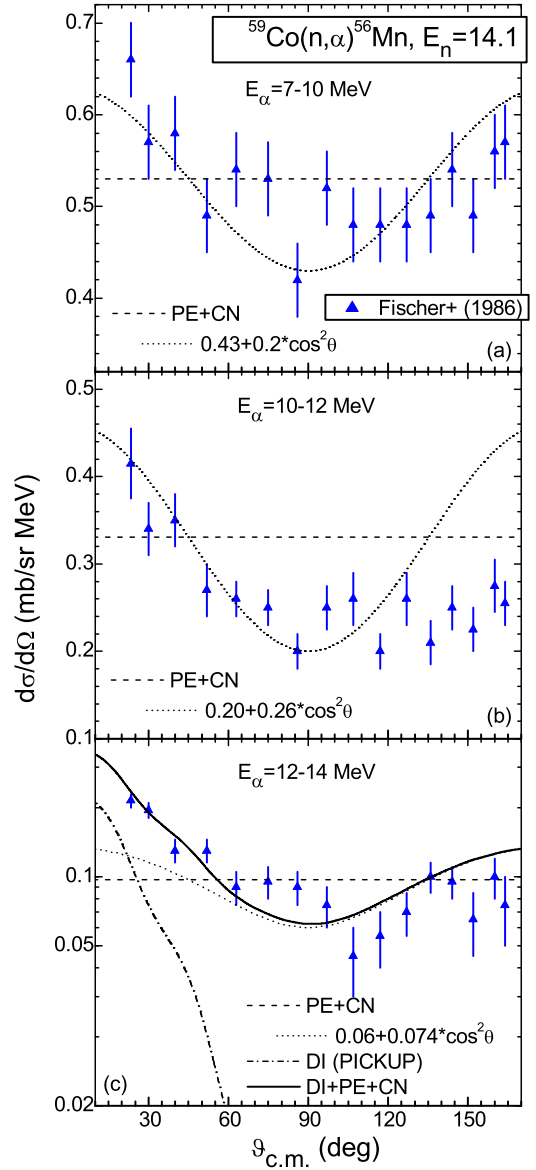}}
\caption{\label{Fig:ad59Cona}  As Fig.~\ref{Fig:ad56Fe141} but for $^{59}$Co target nucleus \cite{rf86} and (a) the first c.m. $\alpha$-energy bin of 7-10 MeV.}
\end{figure}

\subsubsection{$^{59}$Co$(n,\alpha)^{56}$Mn}

Unfortunately, there is no spectroscopic information concerning $^{56}$Mn population through pickup reactions as also $(p,d)$, $(d,t)$, or $(^3He,\alpha)$.
Hence $^{56}$Mn excited states and spectroscopic factors for transitions to neutron-hole states corresponding to the coupling of the $2p_{3/2}$ and $1f_{5/2}$ neutron outside of the magic N=28 shell had to be somehow covered.

Our tentative attempt in this respect concerned the use of neutron spectroscopic factors from the $^{55}$Mn$(d,p)^{56}$Mn reaction analysis \cite{NDS56,comfort} and the spectator protons pair as picked from 1$f_{7/2}$ subshell. 
And lastly, 27 excited states with well-known $J^{\pi}$ and transferred orbital angular momentum \cite{ensdf,ripl3}, until 2.088 MeV excitation energy, were considered in the pickup assessment in the following as well as in the main paper.

A quite similar measurement as the above-mentioned for $^{56}$Fe$(n,\alpha)$ reaction \cite{rf84} was performed on $^{59}$Co target nucleus \cite{rf86} as well. 
Comparison of calculated angular distributions for the c.m. $\alpha$-energy bins of 7-10 MeV, 10-12 MeV, and 12-14 MeV, and the Fischer {\it et al.} data at 14.1 MeV is shown in Fig.~\ref{Fig:ad59Cona}, while comments on the angle-integrated spectrum and total $\alpha$-emission cross section are in the main paper. 

The above--mentioned pickup results were added to the equivalent form $a$+$b$$\cdot$$cos^2$$\theta$ corresponding to the PE+CN isotropic component only for the higher $\alpha$-particle energies (12--14 MeV) in Fig.~\ref{Fig:ad59Cona}(c). 
The lower limit of 2.088 MeV excitation energy for the above-mentioned 27 states of the odd--odd residual nucleus $^{56}$Mn was the origin of this matter, that is yet in agreement with the lowest-lying states being populated through the pickup mechanism.
The anisotropy of the corresponding measured distribution has been better described, with a slight overestimation but yet within $2\sigma$ at backward angles.
The largest distribution, at the $\alpha$-particle energies 7-10 MeV, is also in good agreement with measured data, providing thus a support for the PE+CN present account.

\subsubsection{$^{64}$Zn$(n,\alpha)^{61}$Ni}

\begin{figure} [b]
\resizebox{1.0\columnwidth}{!}{\includegraphics{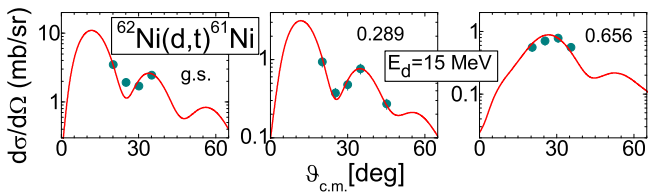}}
\caption{\label{Fig:Ni62dt} As Fig.~\ref{Fig:Ni64pa} but for triton angular distributions of $^{62}$Ni$(d,t)^{61}$Ni pickup transitions \cite{fulmer}.}
\end{figure}

The analysis of the pickup contribution to $^{64}$Zn$(n,\alpha)^{61}$Ni reaction has taken into account the assumption that the spectator protons pair is picked from $2p_{3/2}$ subshell. 
Then, the transferred orbital momenta $L$ to $^{61}$Ni excited states in $(d,t)$ and $(^3He,\alpha)$ reactions that were previously analyzed \cite{ma16,NDS61Ni} were considered as well. 
The spectroscopic factors for the population of the g.s and two excited states in $^{62}$Ni$(d,t)^{61}$Ni reaction were obtained by analysis of triton angular distributions \cite{fulmer} (Fig.~\ref{Fig:Ni62dt}) leading to calculated excitation function in Fig. 6 of Ref. \cite{ma16}. 

\begin{figure} 
\resizebox{1.0\columnwidth}{!}{\includegraphics{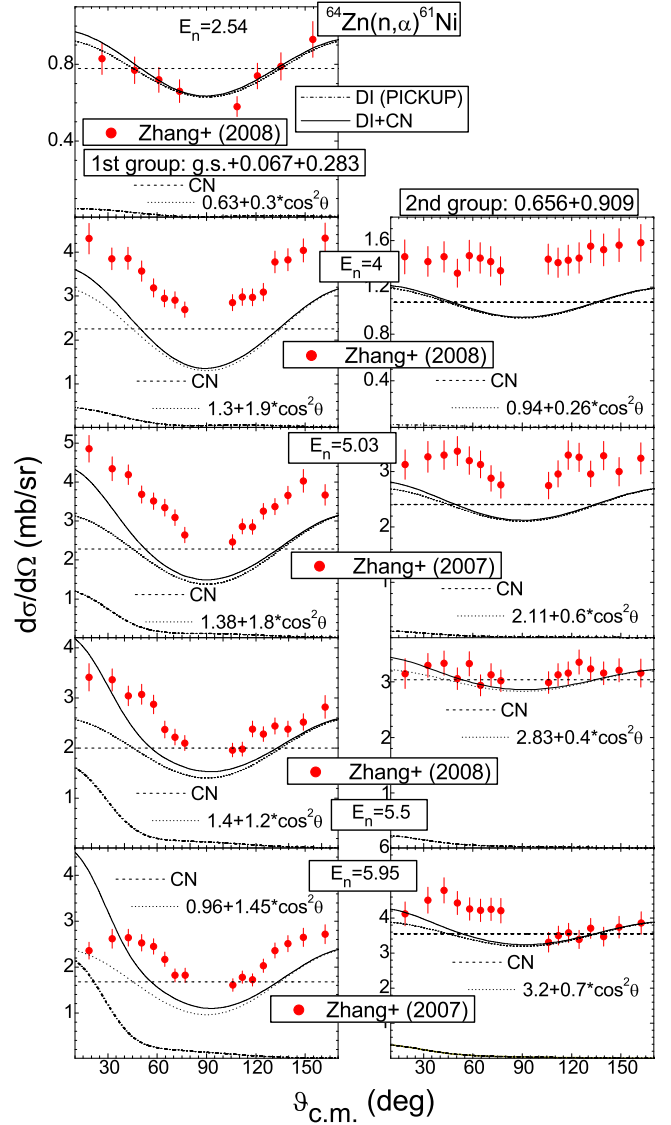}}
\caption{\label{Fig:ad64Znna} As Fig.~\ref{Fig:ad56Fe141} but for $^{64}$Zn target nucleus \cite{gz07,gz08}, neutron energies from 2.54 to 5.95 MeV, and populated g.s. and excited states at 0.067 and 0.283 MeV, within a first group (left side), and 0.656 and 0.909 MeV, in a second group (right side) of $^{61}$Ni.}
\end{figure}

\begin{figure} 
\resizebox{0.7\columnwidth}{!}{\includegraphics{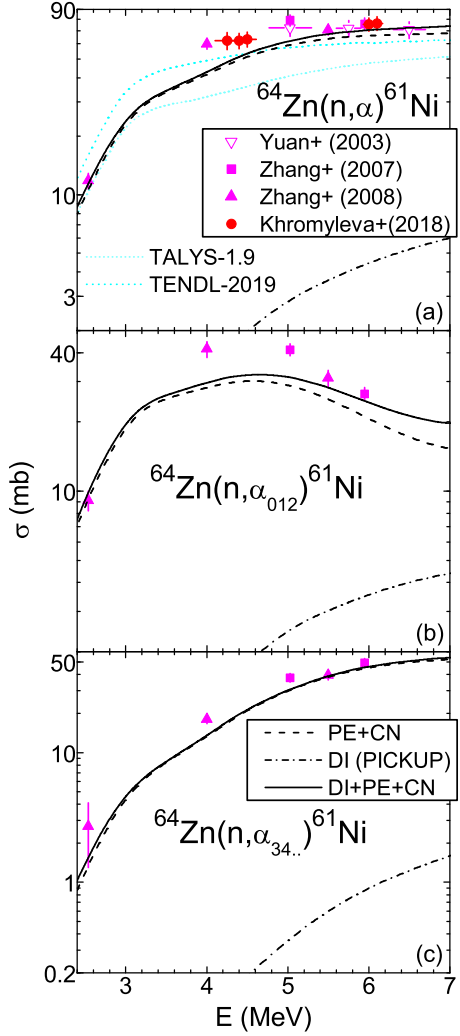}}
\caption{\label{Fig:Zn64na}As Fig.~\ref{Fig:Fe57na} but for (a) $^{64}$Zn$(n,\alpha)$$^{61}$Ni \cite{gz07,gz08,tk18,exfor}, (b) g.s. and excited states at 67, 283, (c) 656, and 909 keV.}
\end{figure}

Next, the spectroscopic factors reported for neutron--removing reaction $^{62}$Ni$(p,d)$ by Schiffer {\it et al.} \cite{schiffer} were used for other 13 excited states until 2.64 MeV excitation energy. 
Most important are however the g.s. and excited states at 0.067 and 0.283 MeV, within a first group, and at 0.656 and 0.909 MeV, within a second group, for which Zhang {\it et al.} \cite{gz07,gz08} measured $\alpha$-particle angular distributions at neutron incident energies from 2.54 to 5.95 MeV. 
The same analysis as for $^{56}$Fe and $^{59}$Co is shown in Fig.~\ref{Fig:ad64Znna} for these data, with major conclusions for the involved reaction mechanisms. 

First, the experimental remark \cite{gz08} of the anisotropic first--group and  almost isotropic second--group angular distributions is confirmed by the pickup contributions to the calculated data. 
It is true even for the incident energies of 4 and 5.03 MeV, where there is an agreement of calculated and measured values only for the data trend, but the same data underestimation as for the total $(n,\alpha)$ cross sections in Fig.~\ref{Fig:Zn64nx}(c). 
On the other hand, the so good agreement at the lowest incident energy of 2.54 MeV is a definite support of the  $\alpha$-particle OMP \cite{va14}, while the earlier one \cite{va94} leads to values that are twice the data.  

This angular--distribution analysis is completed by that of the excitation function for the total $(n,\alpha)$ reaction as well for the above--mentioned first and second group of states (Fig.~\ref{Fig:Zn64na}).
Thus, underestimation for the total $(n,\alpha)$ cross sections at the incident energies of 4 and 5.03 MeV follows the same feature of the first group, with no improvement due to the additional DI contribution. 
However, the good agreement at the lowest incident energy of 2.54 MeV is extended above 5 MeV following the DI inclusion. 
Finally the best agreement proved for the second group at all energies provides a definite support of the  $\alpha$-particle OMP \cite{va14}, the earlier one \cite{va94} leading to values greater than twice the data [Fig.~\ref{Fig:Zn64nx}(c)]. 

A final remark may concern the rather opposite nuclear asymmetries of $^{57}$Fe and $^{64}$Zn. 
Thus the isotopic effect led to so different cross sections, the DI component becoming significant only for the neutron-richer $^{57}$Fe (Fig.~\ref{Fig:Fe57na}).

\section{Conclusions} 
A consistent set of statistical-model input parameters, validated by analysis of various independent data, makes possible the assessment of an optical model potential \cite{va14} also for nucleon-induced $\alpha$-emission within $A$$\sim$60 range.
Particularly,  $\alpha$--emission from $^{55,57,58}$Fe and $^{60}$Co isotopes excited by $(n,\alpha)$ reaction, and $^{62,64,65,66}$Cu and $^{64,65,66,68}$Zn isotopes excited through both $(n,\alpha)$ and $(p,\alpha)$ reactions has been analyzed.
The advantage of rather recent data of low-lying states feeding is essential.
Further consideration of additional reaction channels leading to increase of the $\alpha$-emission cross sections beyond the statistical predictions has also concerned the DI pickup.
The assessment of DI cross sections has been subject to available information on  spectroscopic factors related to populated states, outgoing particle angular distributions, or at least differential cross--section maximum values.

\section*{Acknowledgments}
This work has been partly supported by Autoritatea Nationala pentru Cercetare Stiintifica (Project PN-19060102) and carried out within the framework of the EUROfusion Consortium and has received funding from the Euratom research and training programme 2014-2018 and 2019-2020 under grant agreement No 633053. The views and opinions expressed herein do not necessarily reflect those of the European Commission.

\bibliography{mybibfileA2020}

\end{document}